\documentclass[aps,pre,reprint,superscriptaddress, showpacs]{revtex4-1}

\usepackage{amsmath}
\usepackage{graphicx}
\usepackage{subfigure}
\usepackage{natbib}
\graphicspath{{./Figs/}}  

\begin{document}

\title{Bridging the rheology of granular flows in three regimes}

\author{Sebastian Chialvo}
\affiliation{Chemical and Biological Engineering Department,
		Princeton University, Princeton, NJ 08540, USA}
		
\author{Jin Sun}
\affiliation{Institute for Infrastructure and Environment,
		University of Edinburgh, Edinburgh EH9 3JL, Scotland, UK}
		
\author{Sankaran Sundaresan}
\affiliation{Chemical and Biological Engineering Department,
		Princeton University, Princeton, NJ 08540, USA}
		
\date{\today}


\begin{abstract}
We investigate the rheology of granular materials via molecular
dynamics simulations of homogeneous, simple shear flows of soft,
frictional, noncohesive spheres.  In agreement with previous results for frictionless particles, we observe three flow regimes existing in different domains of particle
volume fraction and shear rate, with all stress data collapsing upon
scaling by powers of the distance to the jamming point.  Though this
jamming point is a function of the interparticle friction coefficient, the
relation between pressure and strain rate at this point is found to be
independent of friction.  We also propose a rheological model that blends the
asymptotic relations in each regime to obtain a general description for
these flows.  Finally, we show that departure from inertial number scalings is a direct result of particle softness, with a dimensionless shear rate characterizing the transition.
\end{abstract}

\maketitle

\section{Introduction}
Flows of granular matter occur in numerous geophysical and industrial processes and, as such, have garnered the attention of researchers for many years.  Early efforts to describe these flows focused on either dilute flows (where kinetic theories~\cite{Garzo1999, Lun1984, Jenkins1985, Johnson1987} apply and which belong to the \textit{inertial} regime) or very dense, slow flows (or \textit{quasi-static} flows, for which plasticity models~\cite{Schaeffer1987, Prevost1985} can be used).  
However, attention has turned recently to the interface between these two regimes in the context of a jamming transition, proposed to occur in granular and other soft matter~\cite{Liu1998}.  Of particular interest are several works that find a critical rheology around this transition in flows of frictionless, soft spheres~\cite{Hatano2008, Otsuki2009b, Nordstrom2010, Seth2008} and disks~\cite{Olsson2007, Tighe2010}.  Furthermore, they find scalings for the mean normal and shear stresses with respect to volume fraction that apply over a wide range of volume fractions and shear rates.  Granular materials, though, are typically considered stiff, frictional materials, and to date there has been little work on identifying a critical rheology~\cite{Campbell2002, Otsuki2011} for such matter despite significant progress in understanding their static jamming behavior~\cite{Zhang2005, Song2008, Majmudar2007, vanHecke2010}.  In this paper, we investigate the rheology of frictional granular matter about the jamming transition and discuss the construction of a rheological model for flows in the quasi-static, inertial, and \textit{intermediate} (i.e.~critical) regimes.

\section{Simulation Methods}
%\textit{Simulation Methods}.~---
We perform computer simulations using a package of the discrete element method (DEM)~\cite{Cundall1979} implemented in the molecular dynamics package LAMMPS~\cite{Plimpton1995}.  In DEM, particles interact only via repulsive, finite-range contact forces.  We employ a spring-dashpot model, for which the normal and tangential forces on a spherical particle $i$ resulting from the contact of two identical spheres $i$ and $j$ are
\begin{align}
	\mathbf{F}_{ij}^{n} &= f(\delta/d) \left[ k_n \delta_{ij} \mathbf{n}_{ij} - \gamma_n m_{\text{eff}} \mathbf{v}_{ij}^n \right] \\
	\mathbf{F}_{ij}^{t} &= f(\delta/d) \left[ -k_t \mathbf{u}_{ij}^t - \gamma_t m_{\text{eff}} \mathbf{v}_{ij}^t \right] \text{,}
	\label{eq:contact_model}
\end{align}
for overlap distance $\delta_{ij}$, particle diameter $d$, spring stiffness constants $k_n$ and $k_t$, viscous damping constants $\gamma_n$ and $\gamma_t$, effective mass $m_{\text{eff}}=m_im_j/(m_i+m_j)$ for particle masses $m_i$ and $m_j$, relative particle velocity components $\mathbf{v}_{ij}^n$ and $\mathbf{v}_{ij}^t$, and elastic shear displacement $\mathbf{u}_{ij}^t$.  A linear spring-dashpot (LSD) model is chosen by setting the function $f(x) = 1$, while a Hertzian model is set by $f(x) = \sqrt{x}$; the LSD model will be used throughout this paper except where noted explicitly.
By Newton's Third Law, particle $j$ experiences the force $\mathbf{F}_{ji}=-\mathbf{F}_{ij}$.  Particle sliding occurs when the Coulomb criterion $|\mathbf{F}_{ij}^t|<\mu |\mathbf{F}_{ij}^n|$ is not satisfied for particle friction coefficient $\mu$.  Additionally, after setting $k_t/k_n=2/7$ and $\gamma_t=0$, we set $\gamma_n$ such that the restitution coefficient $e=$exp$\left(-\gamma_n \pi/\sqrt{4 k_n/m_{\text{eff}}-\gamma_n^2}\right) = 0.7$ in the LSD case.

Using the above contact model, assemblies of about 2000 particles in a periodic box are subjected to homogeneous steady simple shear at a shear rate $\dot{\gamma}$ via the Lees-Edwards boundary condition~\cite{Lees1972}.
The box size, and hence the solids volume fraction $\phi$, are kept constant for each simulation.  The macroscopic stress tensor is calculated as
\begin{equation}
	\boldsymbol{\sigma}=\frac{1}{V}\sum_{i}\left[ \sum_{j \ne i}\frac{1}{2}\mathbf{r}_{ij}\mathbf{F}_{ij} + m_i (\mathbf{v}'_i)(\mathbf{v}'_i)\right]\text{,}
	\label{eq:stress_dem}
\end{equation}
where $V$ is the box volume, $\mathbf{r}_{ij}$ is the center-to-center contact vector from particle $j$ to particle $i$, and $\mathbf{v}'_i$ is the particle velocity relative to its mean streaming velocity; from this result, an ensemble-averaged pressure $p = (\sigma_{xx}+\sigma_{yy}+\sigma_{zz})/3$ and shear stress $\tau = \sigma_{xz}$ can be extracted.  All macroscopic quantities will be presented in dimensionless form, scaled by some combination of the particle diameter $d$, stiffness $k = k_n$, and solid density $\rho_s$.  Since particles are assumed to overlap without deformation, we ensure that the average overlap is small (i.e.~$\delta/d \approx p d/k \lesssim 0.07$).

\section{Flow regimes}
\begin{figure*}
	\centering
	\subfigure[]{
	\includegraphics[width=3.25 in]{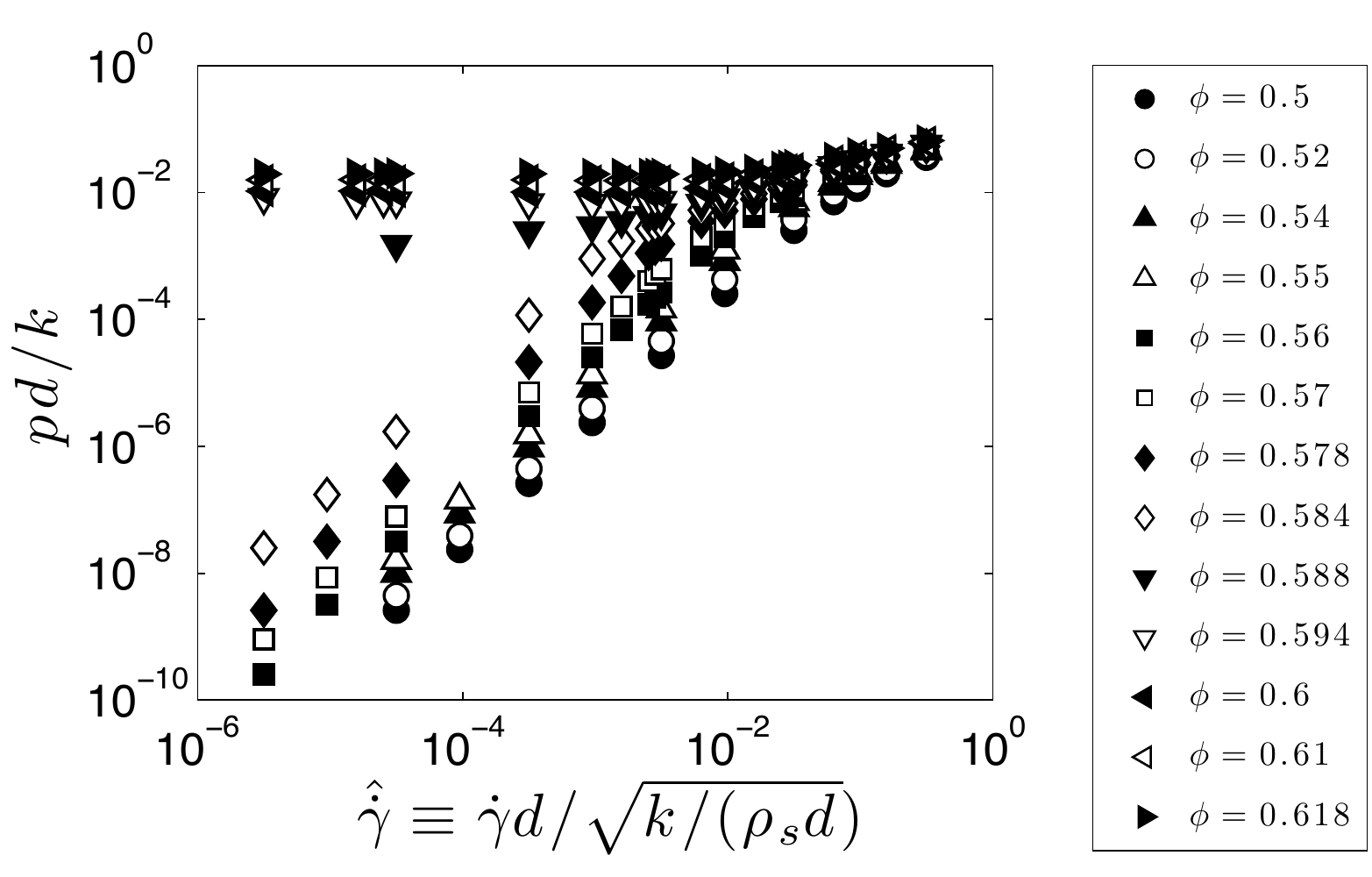}}
	\subfigure[]{
	\includegraphics[width=3.25 in]{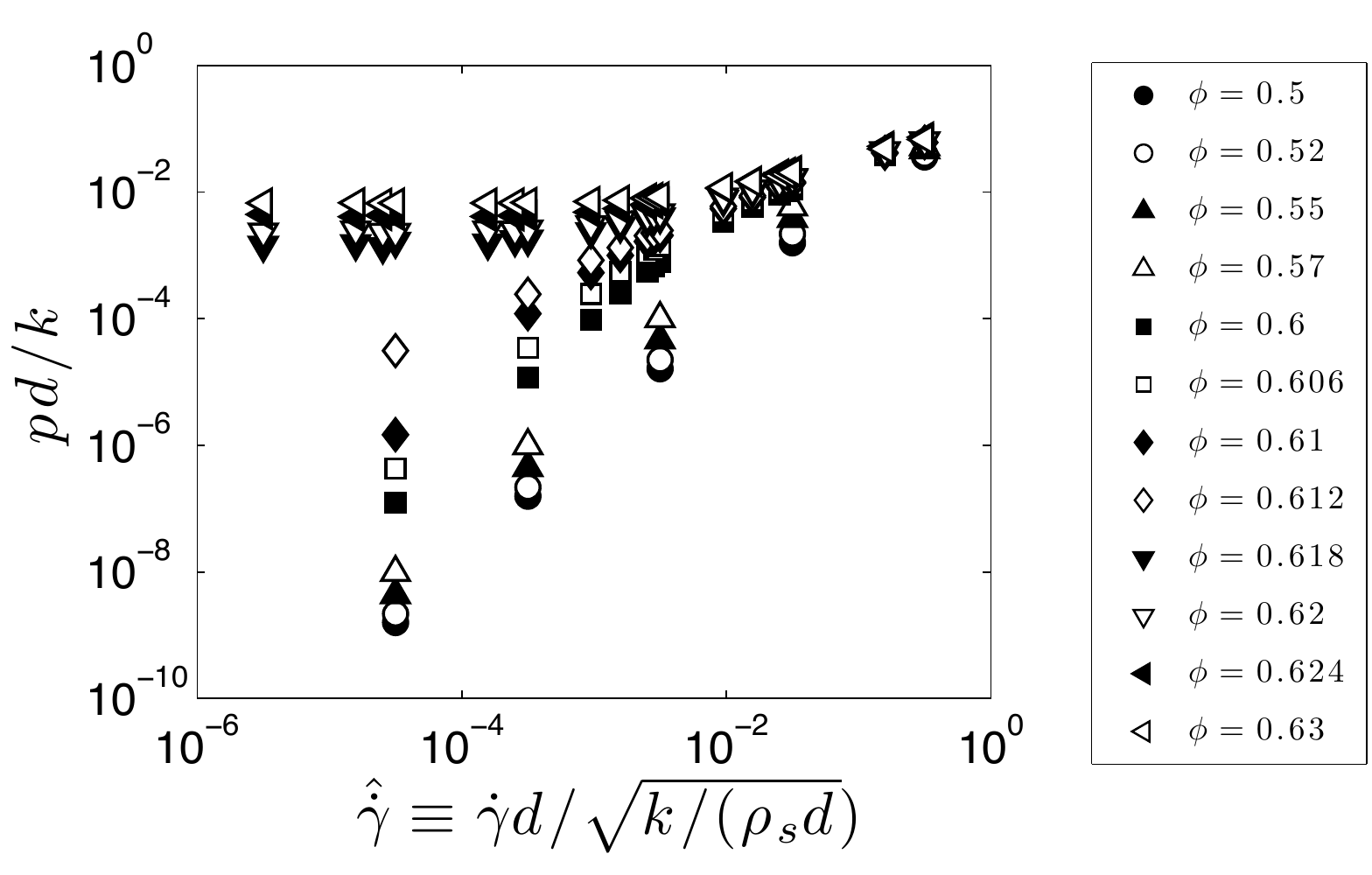}}
%	\subfigure[]{
%	\includegraphics[width=2.75 in]{p_gdot_hertz}}
%	\subfigure[]{
%	\includegraphics[width=2.4 in]{p_gdot_mu01}}
%	\subfigure[]{
%	\includegraphics[width=2.4 in]{p_gdot_mu05}}
%	\subfigure[]{
%	\includegraphics[width=1.9 in]{p_gdot_hertz}}
%	\caption{Dimensionless pressure vs.~dimensionless shear rate for various volume fractions.  Results are shown for (a) $\mu=0.1$ and (b) $\mu=0.5$ with a Hookean spring model, and the $\mu=0.5$ case is repeated for (c) a Hertzian spring model.  In all cases, three flow regimes are observed, each with the scalings $p \sim \dot{\gamma}^m$:  a quasi-static regime with $m = n = 0$, an inertial regime with $m = n = 2$, and an intermediate regime with $m \approx m^*$.  For the Hookean.  Values of $\phi_c(\mu)$ are given in Table~\ref{tab:phi_c}.
\caption{Dimensionless pressure vs.~dimensionless shear rate for various volume fractions with (a) $\mu=0.5$ and (b) $\mu=0.1$.  In both cases, three flow regimes are observed, each with the scalings $p \sim \dot{\gamma}^m$:  a quasi-static regime with $m = 0$, an inertial regime with $m = 2$, and an intermediate regime with $m \approx 1/2$.  At low $\hat{\dot{\gamma}}$, a critical volume fraction $\phi_c$ separates the quasi-static and inertial regimes; values of $\phi_c(\mu)$ are given in Table~\ref{tab:phi_c}.
	\label{fig:stress_gdot}}
\end{figure*}

%\textit{Flow regimes}.~--- 
We performed a series of simple shear simulations over a range of shear rates and volume fractions reaching into all three flow regimes and for several particle friction coefficients between 0 and 1.  Figure~\ref{fig:stress_gdot} shows the scaled pressure $p d/k$ versus the scaled shear rate $\hat{\dot{\gamma}} = \dot{\gamma}d/\sqrt{k/(\rho_{s}d)}$ at various volume fractions for (a) $\mu=0.5$ and (b) $\mu=0.1$.  At low shear rates, there is an observed separatrix occurring at a critical volume fraction $\phi_c$, which we identify as the jamming point; stresses scale quadratically with shear rate below $\phi_c$ but show no rate dependence above it.  These two bands correspond to the inertial and quasi-static regimes, respectively.  As shear rate increases, the quasi-static and inertial isochores approach a shared asymptote characteristic of the critical point in which dependence on the volume fraction vanishes; this region corresponds to the intermediate regime.  Interestingly, the intermediate asymptote appears to be independent of the friction coefficient, in contrast to results at lower shear rates and despite the fact that $\phi_c = \phi_c(\mu)$.
Values of $\phi_c$ for different cases of $\mu$ are presented in Table~\ref{tab:phi_c}.  It should be noted that these critical values inferred from dynamical behavior of sheared systems are unique for each case of $\mu$ and hence may differ from the jamming points of static packings, which are not unique and depend on the compactivity~\cite{Song2008}.
\begin{table}
\caption{Estimates of the critical volume fraction $\phi_c$ for different cases of the interparticle friction coefficient.
	The value of $\phi_c$ for the frictionless case agrees with the experimentally
	determined result of Nordstrom~\cite{Nordstrom2010}.}
\centering
\begin{ruledtabular}
\begin{tabular}{ccccccccccc}
	\textbf{$\mu$}		&&	0.0	&&	0.1	&&	0.3	&&	0.5	&&	1.0	
	\\
	\hline
	\textbf{$\phi_c$}	&&	0.636	&&	0.613	&&	0.596	&&	0.587
						&&	0.581
%	\\
%	\textbf{$\eta_s$}	&&	0.105	&&	0.268	&&	0.357	&&	0.382							&&	0.405	
%
%	\textbf{$Z_c$}	&&	0.636	&&	0.613	&&	0.596	&&	0.587	&&	0.581	
\end{tabular}
\end{ruledtabular}
\label{tab:phi_c}
\end{table}

\begin{figure}
	\centering
	\includegraphics[width=3 in]{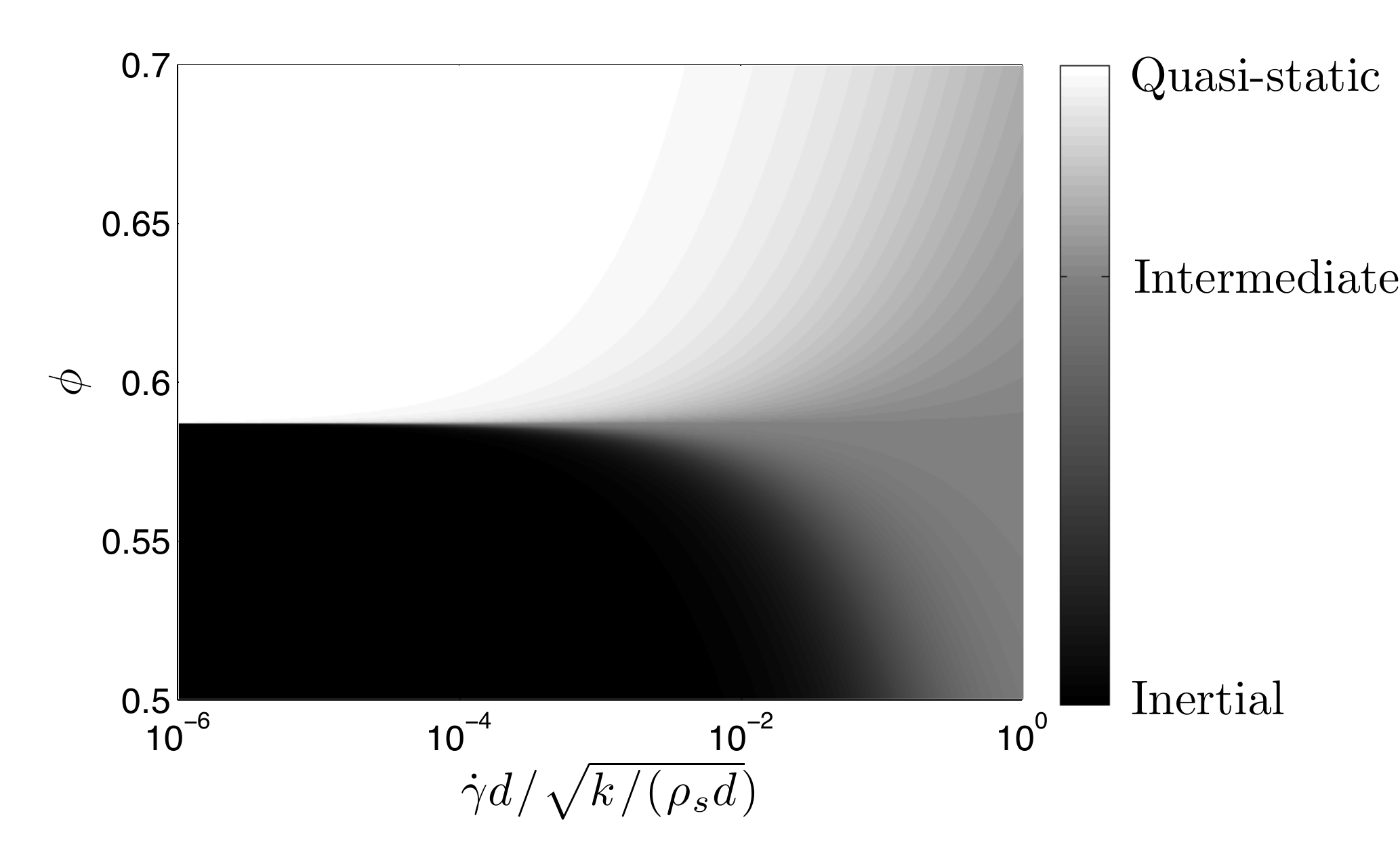}
	\caption{Regime map for $\mu=0.5$, with volume fraction vs. dimensionless shear rate.  The intermediate regime occurs only at $\phi_c$ in the limit $\hat{\dot{\gamma}} \to 0$ but emerges from this point to encompass all volume fractions as $\hat{\dot{\gamma}} \to \infty$.
	\label{fig:regime_map}}
\end{figure}

A better understanding of the regime transitions can be gained by constructing a regime map, or ``phase diagram," from the slopes of the curves in Figure~\ref{fig:stress_gdot}.  Such a map is shown in Figure~\ref{fig:regime_map}.  The intermediate regime is observed to lie in a window centered around $\phi = \phi_c$, and the width of this window is dependent on the value of the dimensionless shear rate.
This feature has important implications for the modeling of dense granular flows.  The large stiffness of granular materials such as sand or glass beads has been used to justify the modeling of granular particles as (infinitely) hard spheres.  For such particles, dimensional analysis requires the traditional Bagnold scaling of the stresses (i.e.~$p, \tau \sim \dot{\gamma}^2$), thereby rendering the intermediate and quasi-static regimes impossible.  This picture is consistent with the vanishing of the intermediate regime observed in Figure~\ref{fig:regime_map} as $k \to \infty$.  However, real granular materials do nevertheless have a finite stiffness.  Therefore, in the context of building a general rheological model for granular flows, it is preferable to choose a framework that include all three regimes.

Another important observation from Figure~\ref{fig:regime_map} is the smoothness of the transitions between the regimes.  This feature suggests that purely quasi-static, inertial, or intermediate flow is achieved only in certain limits.  As $\hat{\dot{\gamma}} \to 0$, we see quasi-static flow for $\phi > \phi_c$, inertial flow for $\phi < \phi_c$, and intermediate flow at $\phi = \phi_c$.  We also see intermediate flow as $\hat{\dot{\gamma}} \to \infty$ for all volume fractions over the wide range examined in this study.  The smooth transitions also suggest that the rheology at a particular $(\dot{\gamma},\phi)$ is a composite of contributions from low-$\hat{\dot{\gamma}}$ and high-$\hat{\dot{\gamma}}$ behaviors, and this notion will play a large role in our construction of a rheological model.

\section{Critical volume fraction $\phi_c$} \label{sec:phi_c}
\begin{figure*}
	\centering
	\subfigure[]{
	\includegraphics[width=3 in]{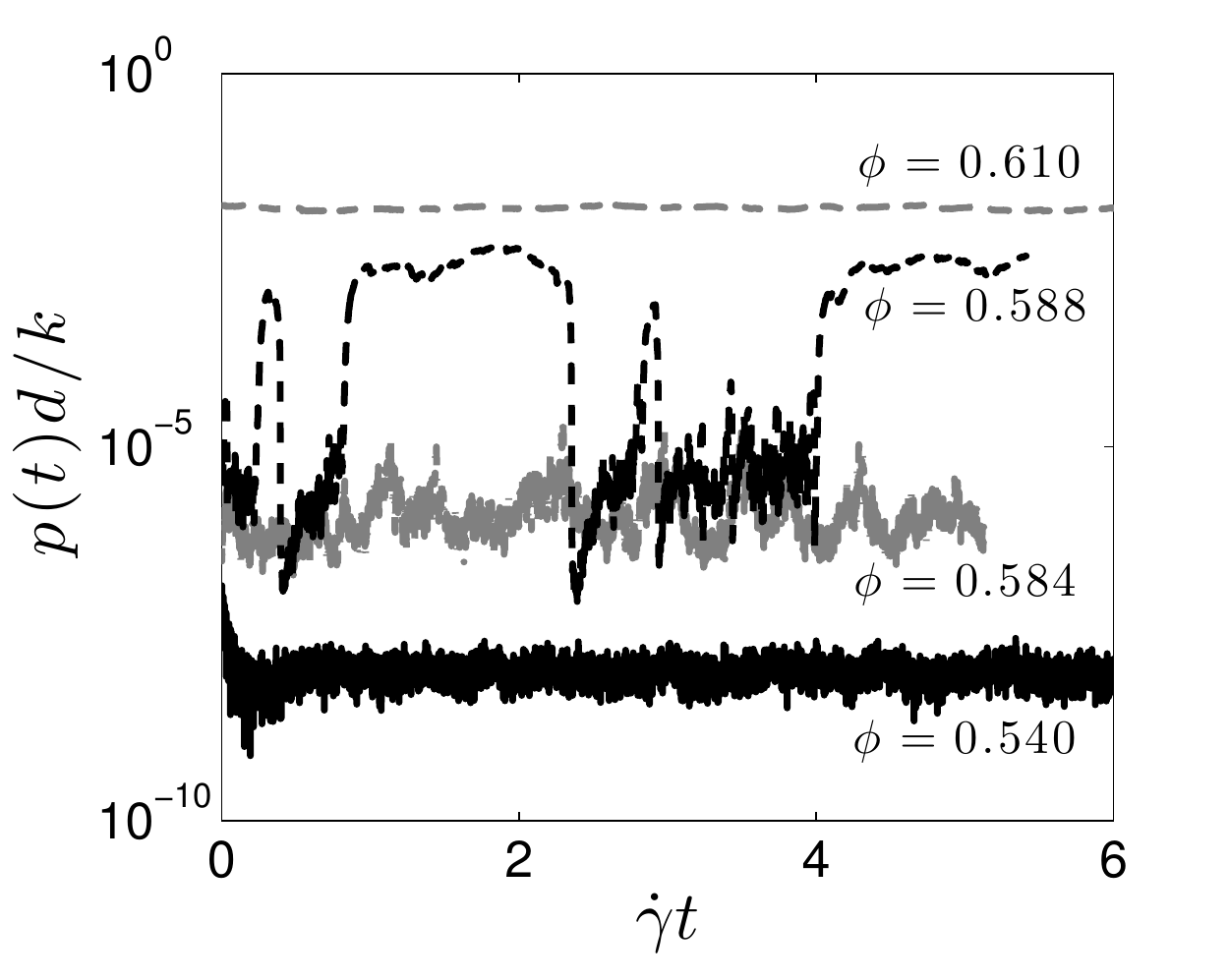}}
	\subfigure[]{
	\includegraphics[width=3 in]{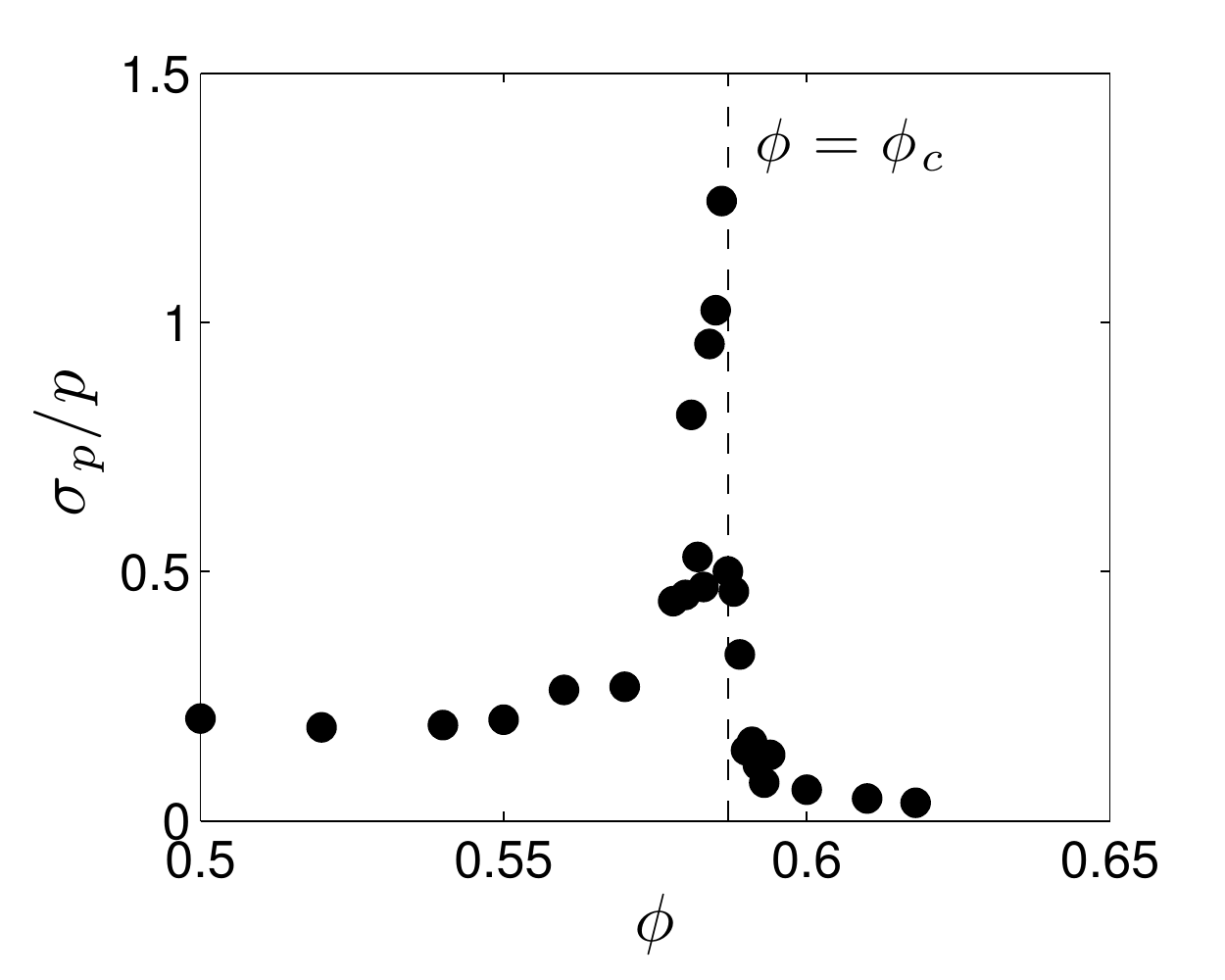}}
	\subfigure[]{
	\includegraphics[width=3 in]{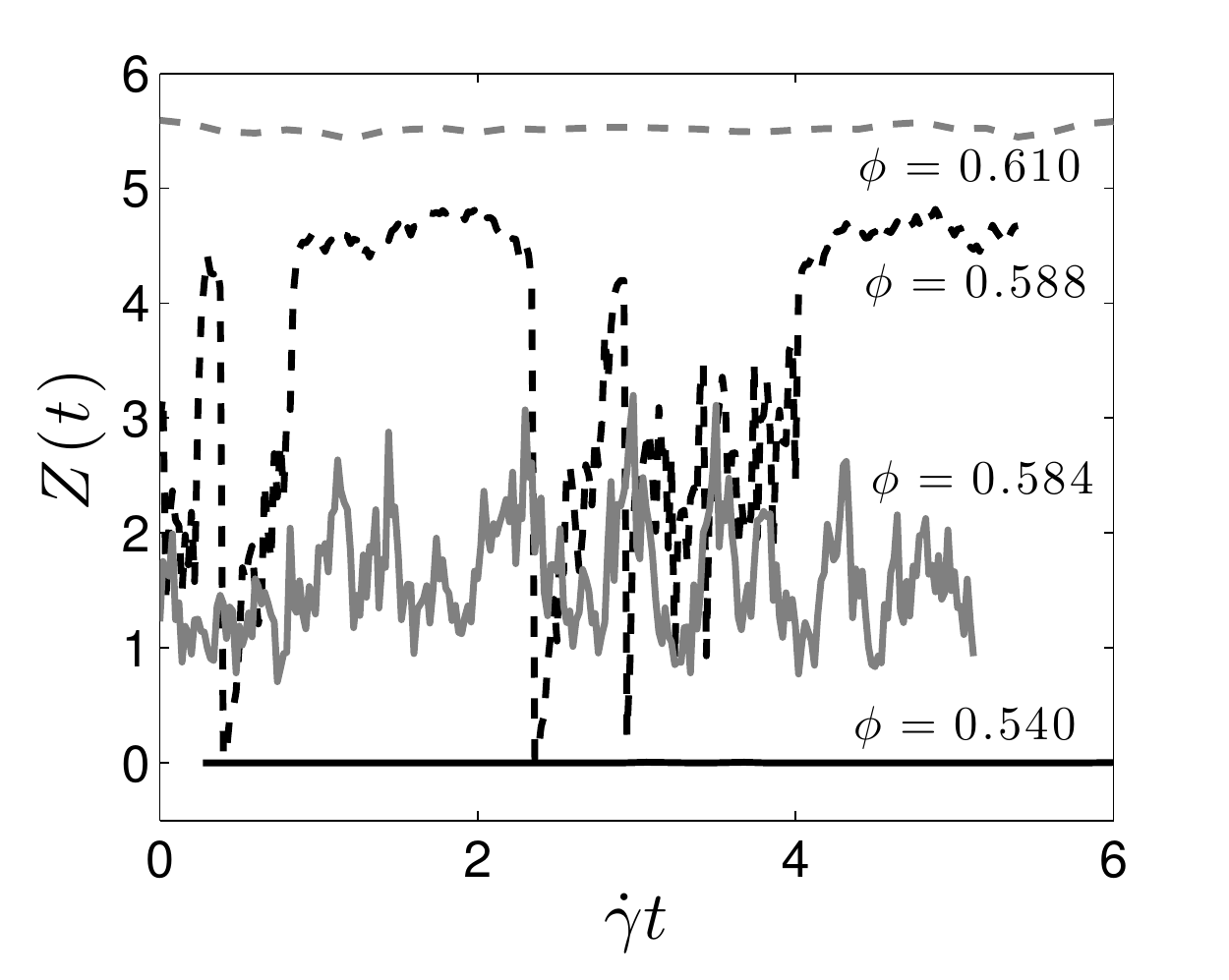}}
	\subfigure[]{
	\includegraphics[width=3 in]{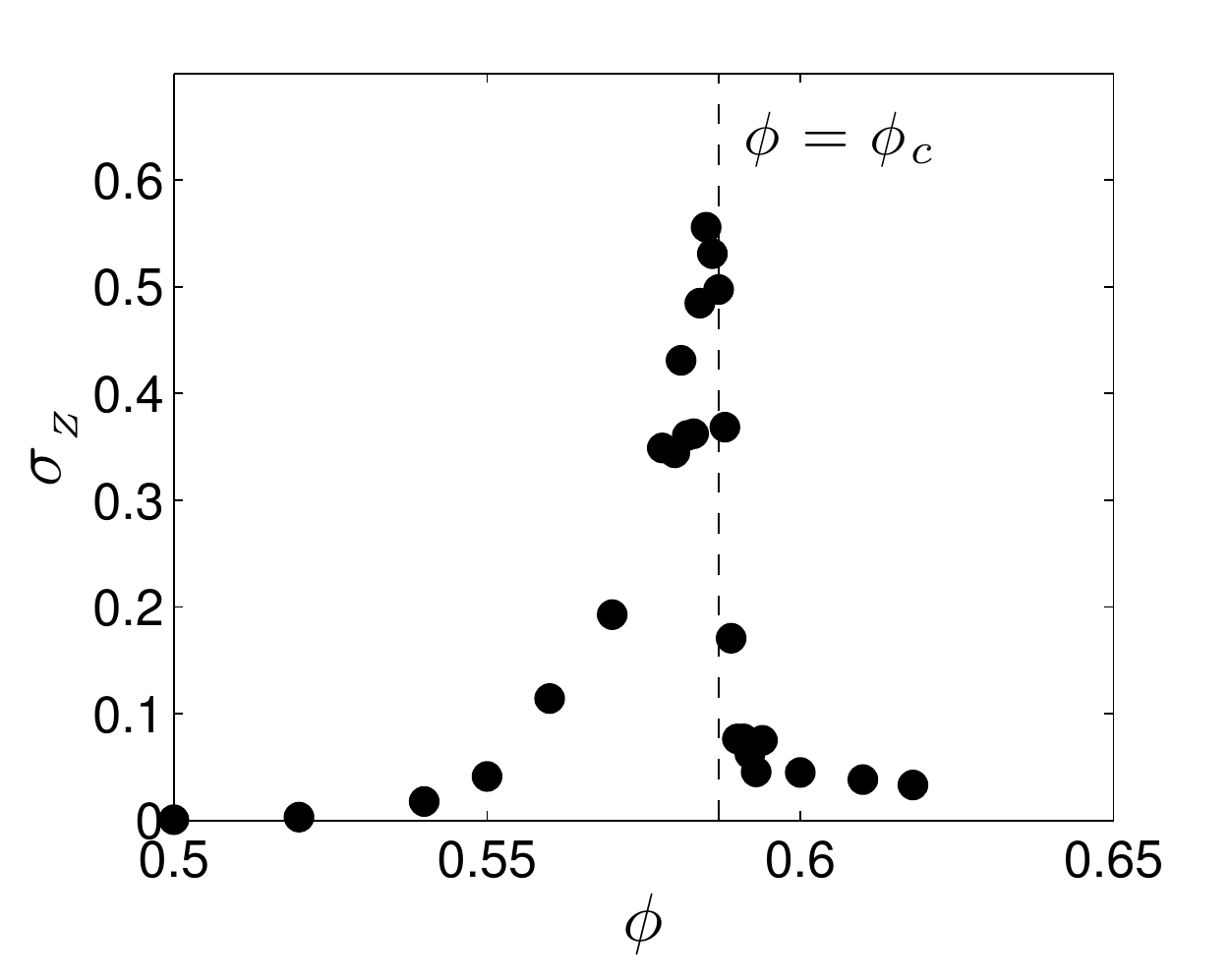}}
%	\includegraphics[width=3.25 in]{p_t}}
%	\subfigure[]{
%	\includegraphics[width=3.25 in]{pstd}}
%	\subfigure[]{
%	\includegraphics[width=3.25 in]{Z_t}}
%	\subfigure[]{
%	\includegraphics[width=3.25 in]{p_Z}}
	\caption{Characteristics of the critical point for $\mu = 0.5$ and $\hat{\dot{\gamma}} = 3.2\times 10^{-5}$.  (a) Pressure fluctuations in time are observed to become larger near the critical point.  (b) The standard deviation of pressure, scaled by the mean pressure, exhibits a spike at $\phi_c$.  (c-d) The coordination number fluctuations are similar to those of the pressure in terms of both dynamics and $\phi$-dependence.
	\label{fig:fluct}}
\end{figure*}

%\textit{Critical volume fraction $\phi_c$}.~---
Because $\phi_c$ plays such an important role in governing the rheology in each of the three flow regimes, accurate estimation of its value for each case of $\mu$ is required for the construction of a valid rheological model.  However, this task is made difficult by fluctuations of our measurements in time $t$.  We observe a propensity for assemblies near $\phi_c$ to form and break force chains intermittently during the shearing process, resulting in stress fluctuations of several orders of magnitude as seen in Figure~\ref{fig:fluct}a.  Though fluctuations occur at all volume fractions, their size relative to the mean is markedly large near the critical point.  In Figure~\ref{fig:fluct}b the standard deviation $\sigma_p \equiv \sqrt{\langle p^2(t) \rangle - \langle p(t) \rangle ^2}$ of the pressure, when scaled by the time-averaged pressure $p$, exhibits a spike centered slightly under $\phi_c$.  This phenomenon increases the potential error in the time-averaged stress values near the critical point, thereby limiting the precision of our $\phi_c$ estimates to within about $\pm 0.001$.

Additionally, though $\phi_c$ is certainly an important quantity, it is not necessarily the only or even the most influential parameter describing the jamming transition.  The fact that stress can vary significantly at a constant volume fraction indicates that, while $\phi$ is a useful predictor of time-averaged stresses, other state variables may be more suitable for predicting instantaneous stresses.  This quality has been observed previously with the coordination number $Z(t)$, for example, which was shown to exhibit a one-to-one correspondence with $p(t)$ in the quasi-static regime~\cite{Sun2011}.  Indeed, we observe that $p(t)$ and $Z(t)$ exhibit the same qualitative evolution in time (Figure~\ref{fig:fluct}c) and a similar $\phi$-dependence in their fluctuations (Figure~\ref{fig:fluct}d).  Here we define $Z(t) \equiv 2N_c(t)/N$ for $N_c$ contacts occurring in the $N$-particle assembly.  The $p(t)$-$Z(t)$ relationship suggests that the critical point is better defined by some critical coordination number $Z_c$.  However, because our goal is the construction of a \textit{steady-state} rheological model, it is convenient to ignore all dynamics and assume that $Z_c$ and $\phi_c$ correspond to the same conditions.  We therefore proceed with $\phi_c$ as the definition of the critical point for our model.
%We note values of $Z_c$ for different cases of $\mu$ in Table~\ref{tab:phi_c} but nevertheless proceed with $\phi_c$ as the definition of the critical point for our model.
%
%To this end, we analyze the variation of pressure with volume fraction in the vicinity of $\phi_c$.  
%We observe a propensity for assemblies near $\phi_c$ to form and break force chains intermittently during the shearing process, resulting in stress fluctuations of several orders of magnitude.  This phenomenon limits the precision of the $\phi_c$ estimates to within about $\pm 0.001$.

\begin{figure*}
	\centering
	\subfigure[]{
	\includegraphics[width=2.2 in]{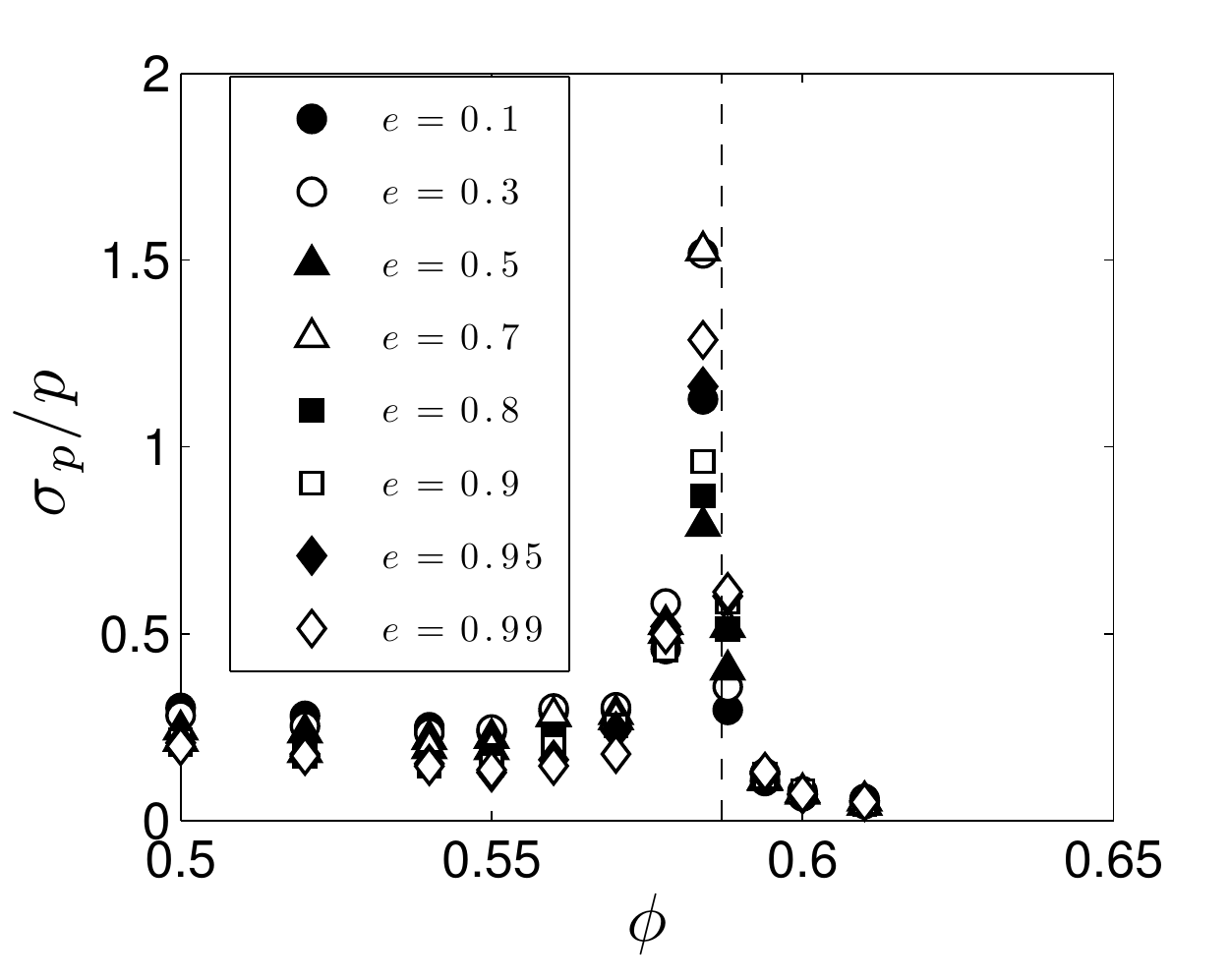}}
	\subfigure[]{
	\includegraphics[width=2.2 in]{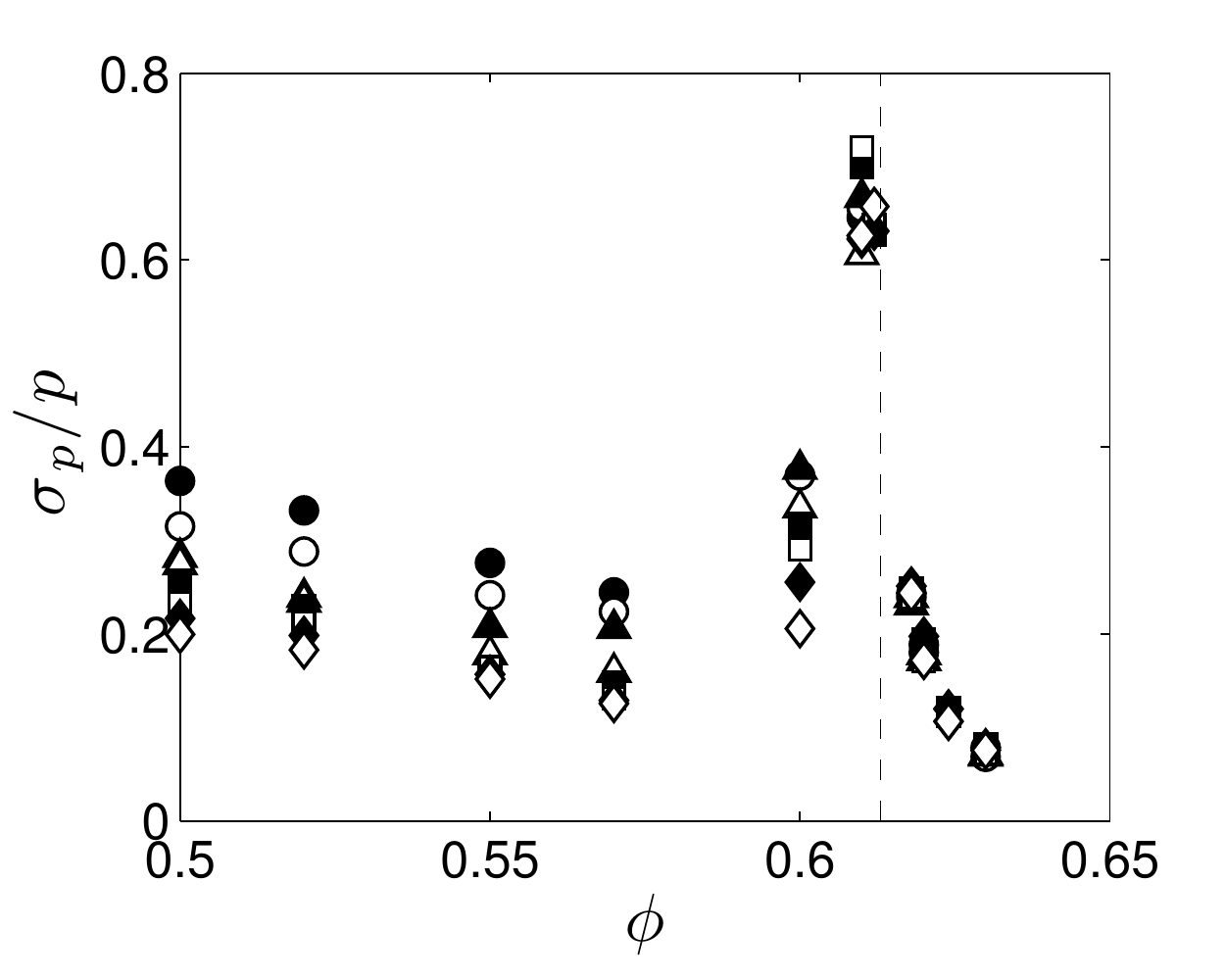}}
	\subfigure[]{
	\includegraphics[width=2.2 in]{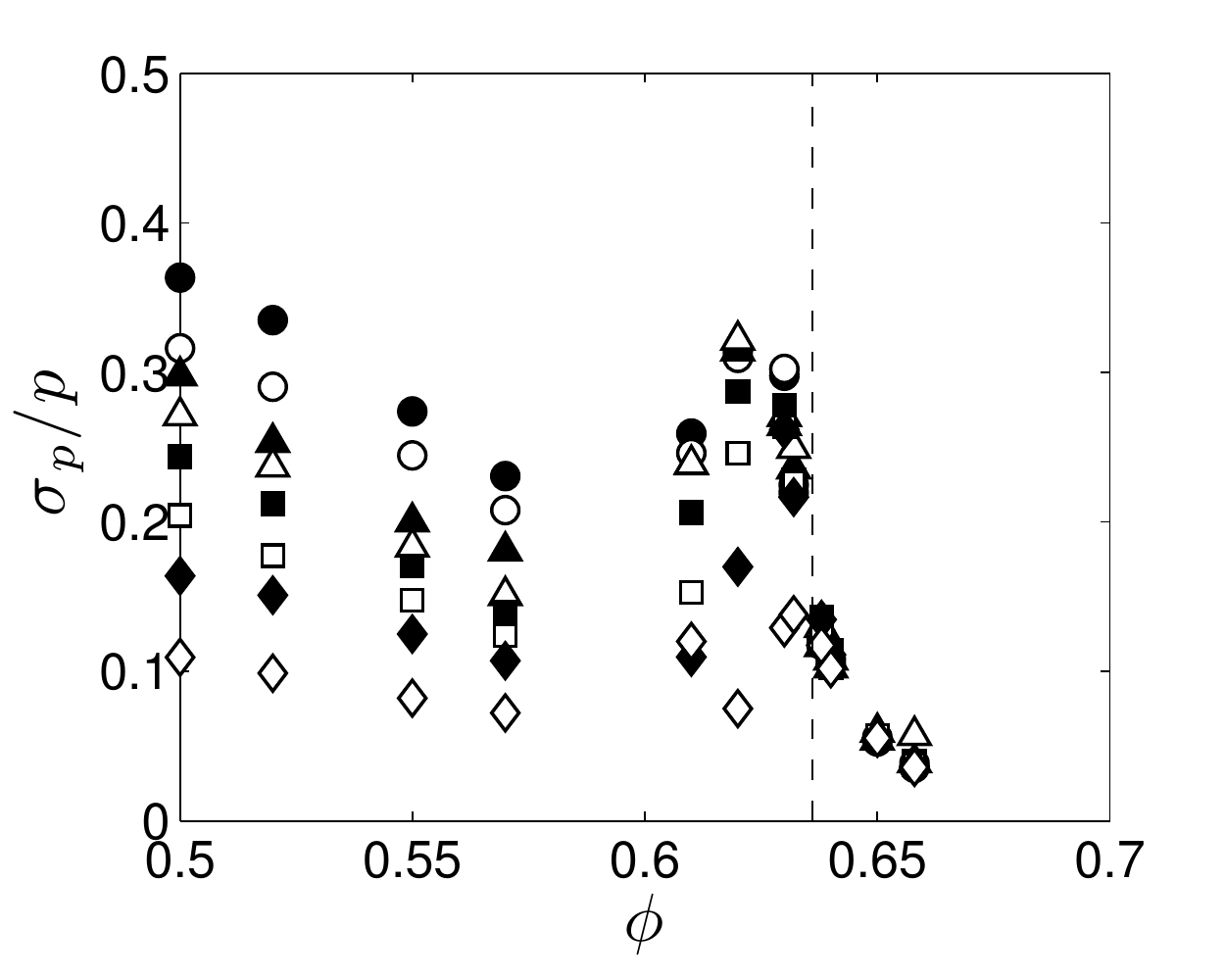}}
	\caption{Effect of changing the restitution coefficient on the pressure fluctuations for $\hat{\dot{\gamma}} = 3.2\times 10^{-5}$.  Dotted lines demarcate the critical point $\phi_c$.  For (a) $\mu = 0.5$ and (b) $\mu = 0.1$, the location of the spike in pressure fluctuations is independent of $e$.  (c) Even for the frictionless case, there is little evidence to suggest $\phi_c = \phi_c(e)$.
	\label{fig:e_dep}}
\end{figure*}

In addition to being a function of $\mu$, the critical point has also been proposed to change with the restitution coefficient $e$~\cite{Kumaran2009a}, and such a $\phi_c(e)$ has been used in a kinetic theory for frictional particles~\cite{Jenkins2010}.  However, our DEM results do not support this conclusion, especially for frictional particles.  As seen in Figure~\ref{fig:e_dep}a-b for $\mu = 0.5$ and $\mu = 0.1$, the spike in the pressure fluctuations occurs at the same volume fraction for a given $\mu$ regardless of the value of $e$, suggesting that $\phi_c = \phi_c(\mu)$ only.  Even for the frictionless case (Figure~\ref{fig:e_dep}c), where fluctuations tend occur over a wider range of volume fractions, there is no clear trend in the peak towards lower $\phi$.  One possible reason for the discrepancy is the methods used for determining $\phi_c$.  Because hard-sphere codes, used in Ref.~\cite{Kumaran2009a}, prohibit particle overlaps, they are unable to simulate sheared particle systems near or above $\phi_c$~\cite{PoschelSchwager}.  This shortcoming limits the performable simulations to one side of $\phi_c$, thus requiring the critical point to be estimated via extrapolation.
Furthermore, while hard-sphere methods treat collisions as binary interactions, entrance into the quasi-static regime coincides with the development of multi-body interactions that persist even in the hard-sphere limit~\cite{Mitarai2003}.  This conflict may render even-driven algorithms less accurate at resolving collisions upon approaching $\phi_c$ and perhaps result in an erroneous estimation of the value of $\phi_c$.  Soft-sphere DEM, on the other hand, enables us to resolve multi-body contacts and simulate shear flows at any volume fraction on either side of $\phi_c$, thereby allowing us to \textit{interpolate} the value of $\phi_c$.  For these reasons, we expect the latter approach to provide more accurate $\phi_c$ estimates.

%With soft-sphere methods, however, there is no such restriction on the volume fraction.  Hence, simulations can be performed on both sides of $\phi_c$, which can then be inferred by interpolation.  Additionally, though hard-sphere methods treat collisions as binary interactions, sphere packings are known to exhibit a significant increase in the coordination number near $\phi_c$~\cite{}.  These multi-body interactions are characterized by a broad distribution of contact lifetimes~\cite{Silbert2007}.  We expect the latter approach to provide more accurate $\phi_c$ estimates.

%While $\phi_c$ does not change with $e$, the radial distribution function $g_0(\phi)$.

\section{Pressure scalings and regime blending}
\begin{figure*}
	\centering
	\subfigure[]{
	\includegraphics[width=2.5 in]{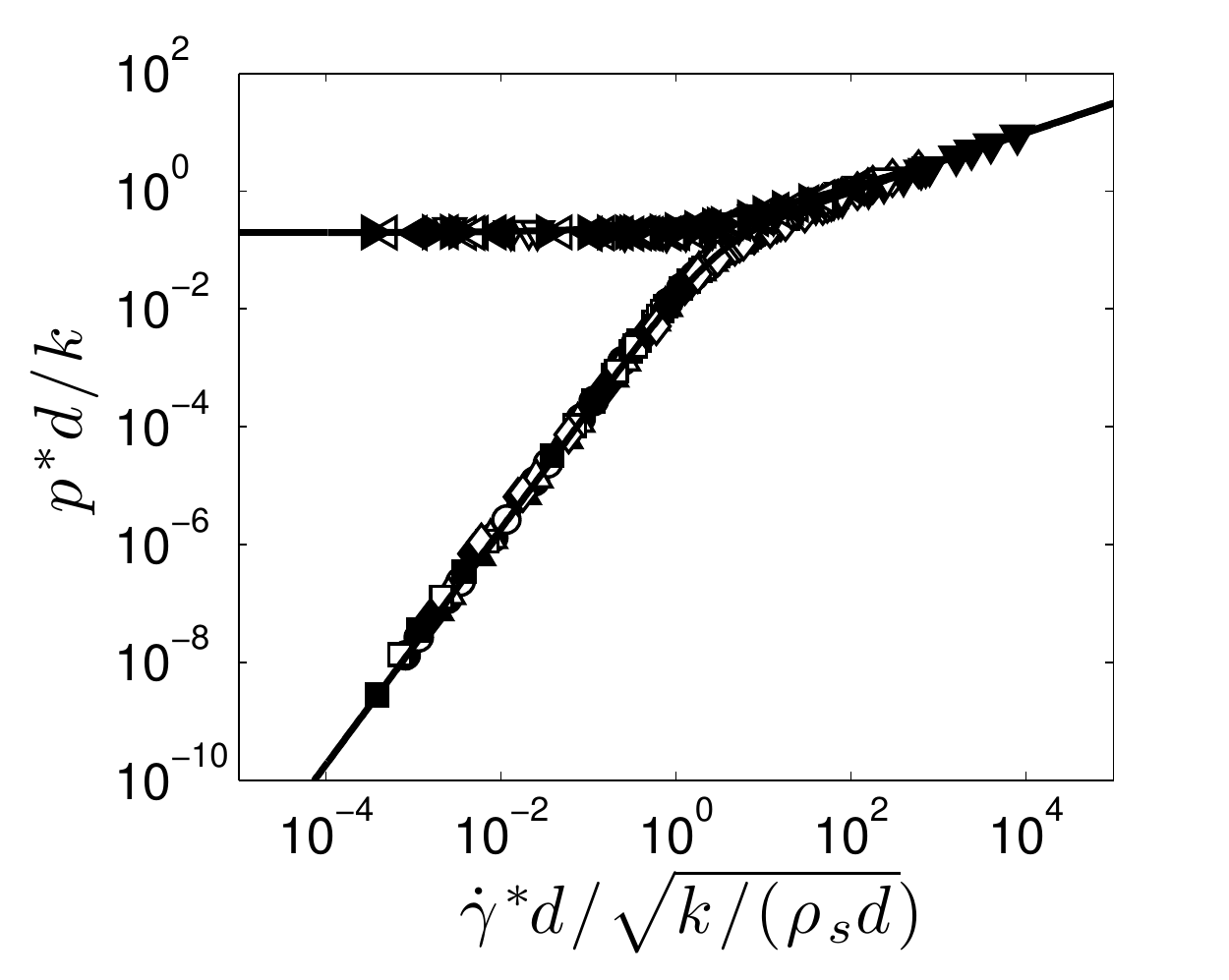}}
	\subfigure[]{
	\includegraphics[width=2.5 in]{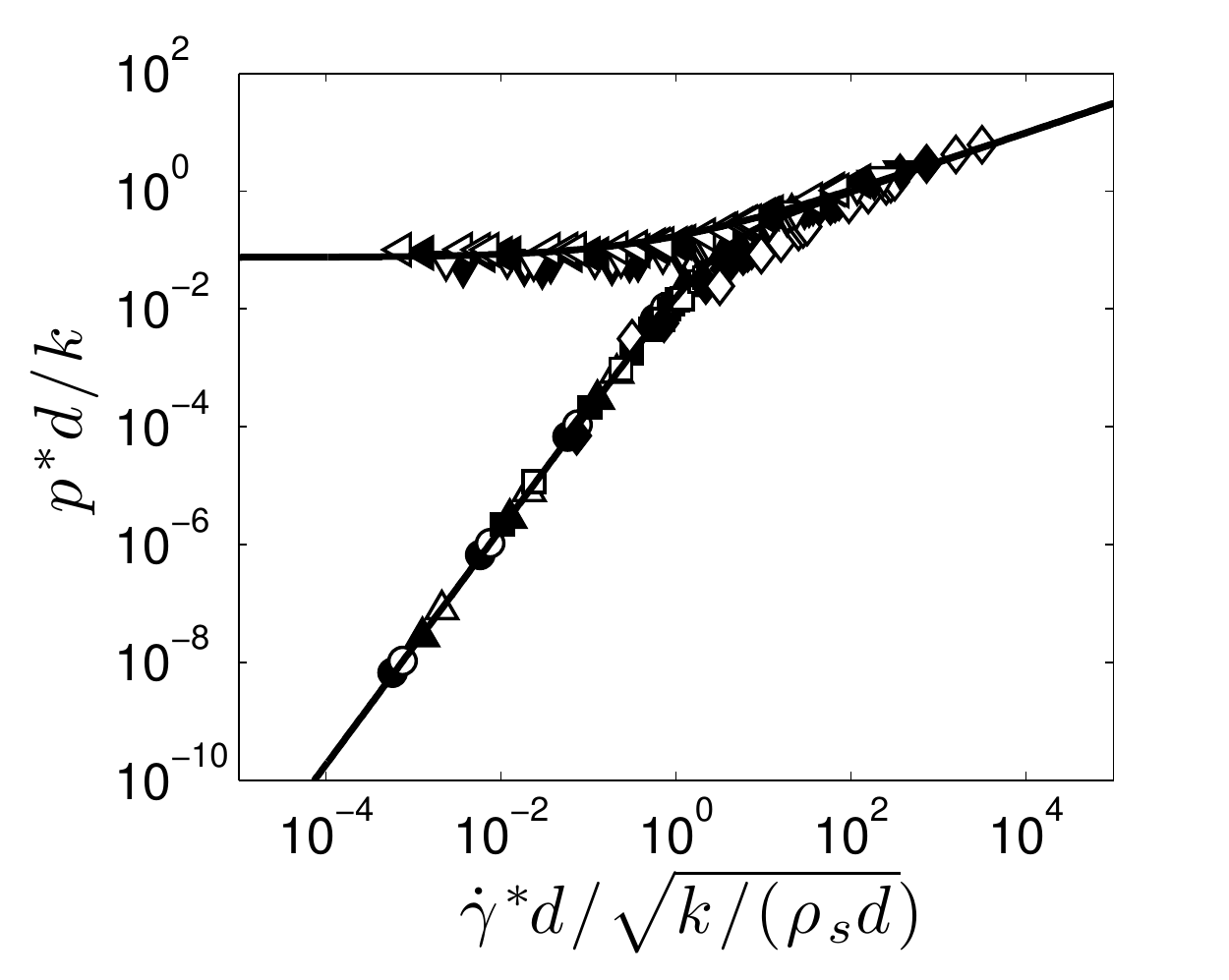}}
	\caption{Collapse of pressure vs.~shear rate curves from Figure~\ref{fig:stress_gdot} for (a) $\mu=0.5$ and (b) $\mu=0.1$.  In both cases, the pressure is scaled as $p^*=p/|\phi-\phi_c|^{2/3}$ and the shear rate as $\dot{\gamma}^*=\dot{\gamma}/|\phi-\phi_c|^{4/3}$.  A simple blending function (solid lines) captures regime asymptotes and transitions.
	\label{fig:hatano}}
\end{figure*}

\begin{figure}
	\centering
	\includegraphics[width=2.5 in]{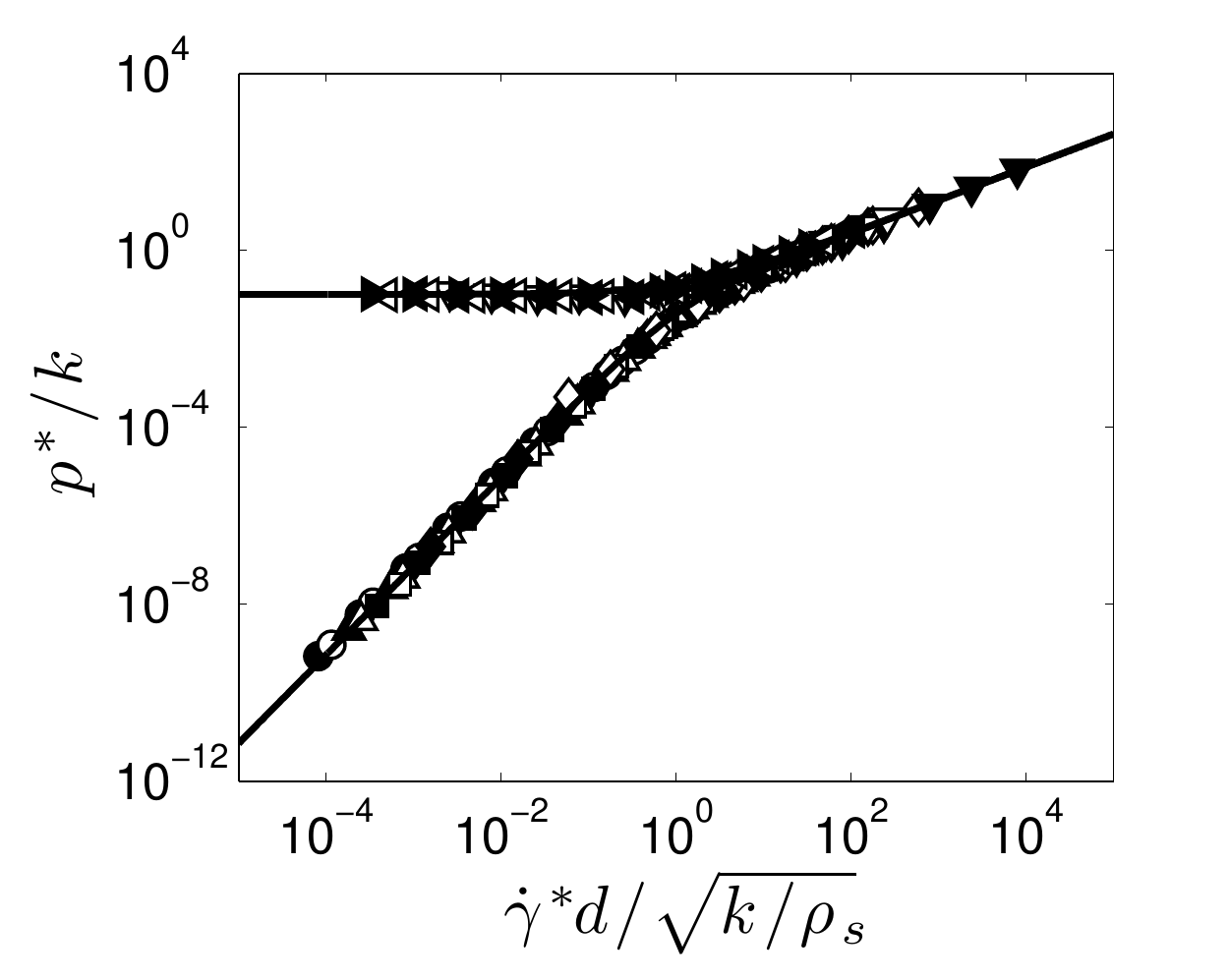}
	\caption{Collapse of pressure vs.~shear rate curves for Hertzian particles with $\mu=0.5$.  The volume fractions (and legend) are the same as from Figure~\ref{fig:stress_gdot}(a).  Here, $p^*=p/|\phi-\phi_c|^{1}$, $\dot{\gamma}^*=\dot{\gamma}/|\phi-\phi_c|^{4/3}$, and $m^* \approx 3/4$.  Regime asymptotes and transitions are captured by the same blending function (solid lines) as in the Hookean case.
	\label{fig:hertz}}
\end{figure}

%\textit{Pressure scalings}.~---
It has been demonstrated in experimental~\cite{Nordstrom2010} and computational~\cite{Hatano2008, Otsuki2009b, Olsson2007} studies of frictionless particles that stress data will collapse onto two curves (one above $\phi_c$ and one below) upon scaling the stresses and shear rate by powers of $|\phi-\phi_c|$, the distance to jamming.  This idea is consistent with several models of the radial distribution function, used in kinetic theories for the inertial regime, that diverge at close packing~\cite{Torquato1995, Lun1986, Ogawa1980}.  Such a collapse can be achieved for frictional particles as well, as shown in Figure~\ref{fig:hatano}, with
\begin{eqnarray}
	p^*=p/|\phi-\phi_c|^{a} & \hspace{0.25 in} & \dot{\gamma}_p^*=\dot{\gamma}/|\phi-\phi_c|^{b}
%	\\
%	\tau^*=\tau/|\phi-\phi_c|^{a_2} & \hspace{0.25 in} & \dot{\gamma}_{\tau}^*=\dot{\gamma}/|\phi-\phi_c|^{b_2}
\end{eqnarray}
and constitutive exponents $a$ and $b$.  This result for frictional disks was also found independently in Ref.~\cite{Otsuki2011}.  From the collapse it is clear that an asymptotic power-law relationship between stress and shear rate exists for each flow regime $j$, and we can write the form of each asymptote as
\begin{align}
	\frac{p_j}{|\phi-\phi_c|^{a}} &\sim \left[ \frac{\dot{\gamma}}{|\phi-\phi_c|^{b}} \right]^{m_j}
	\label{eq:asymptote_p}
%	\\
%	\frac{\tau_j}{|\phi-\phi_c|^{a_2}} &\sim \left( \frac{\dot{\gamma}}{|\phi-\phi_c|^{b_2}} \right)^{n_j}
%	\label{eq:asymptote_tau}
\end{align}
where $m_{\text{QS}}=0$, $m_{\text{Inert}}=2$, and $m_{\text{Int}}=m^*$.  The exponents $a$ and $b$ can be fitted from the DEM data, but the values are sensitive to the choice of $\phi_c$ used~\cite{Otsuki2009b} and hence should be chosen with care.  Our inertial regime data suggest that $p_{\text{Inert}} \sim |\phi-\phi_c|^{-2}$, which is consistent with previous results~\cite{daCruz2005}; quasi-static regime data reveal that $p_{\text{QS}} \sim |\phi-\phi_c|^{2/3}$; and, as noted earlier, $p_{\text{Int}} \sim |\phi-\phi_c|^{0}$. These trends lead us to set $a = 2/3$, $b=4/3$, and $m^*=1/2$.  The $m^*$ value is consistent with our fits of the intermediate asymptote (Table~\ref{tab:phi_c}) and with experimental results~\cite{Nordstrom2010, Seth2008}, and it is similar to other values proposed for frictionless particles using the linear spring-dashpot model~\cite{Hatano2008, Otsuki2009b}.
%The difference between the value of $a$ here and that in Ref.~\cite{Sun2011} ($a=1$) is likely due to the different $\phi_c$ values used.  Using their value of $a$ still yields a reasonable   
The value of $a$ used in Ref.~\cite{Sun2011} ($a=1$), though different, still yields a decent collapse.  However, in that work, $\phi_c$ is determined by extrapolation from the quasi-static regime, while here we interpolate it from quasi-static and inertial regime data and furthermore verify it with stress fluctuation data, as described in Section~\ref{sec:phi_c}; hence we believe our current $\phi_c$ values and the resulting $a$ value to be more accurate.
We also point out that the above scaling exponents depend on the contact model used~\cite{Hatano2008, Otsuki2009a}.  Based on a small set of simple shear simulations with a Hertzian contact model, we observe the values of $a \approx 1$ and $m^* \approx 3/4$ to be larger than in the LSD case by a factor of $3/2$, which is consistent with previous results for static, jammed systems~\cite{Zhang2005}.  The value of $b$, however, remains the same for both contact models; note that in both cases $a = b m^*$ in order to satisfy the functional forms implied by the collapse.  The resulting collapse for the Hertzian particles is shown in Figure~\ref{fig:hertz}.

%\section{Regime blending}
%\textit{Regime blending}.~---
Though the individual regime limits can be described using Eq.~\ref{eq:asymptote_p}, the transitions between them have yet to be modeled.  To this end, we employ a blending function $B$ of the form
\begin{equation}
	B(y_1,y_2)=(y_1^w + y_2^w)^{1/w}
	\label{eq:blending}
\end{equation}
with $w>0$ yielding an additive blend for the quasi-static-to-intermediate transition and $w<0$ providing a harmonic blend for the inertial-to-intermediate transition.  Figure~\ref{fig:hatano} demonstrates the use of Eq.~\ref{eq:blending} with the asymptotic forms of Eq.~\ref{eq:asymptote_p} and $w=\pm 1$.
The blended model is able to capture the pressure behavior continuously in shear rate for all three regime limits as well as the transitions; moreover, it does so without defining the stresses in piecewise fashion over arbitrary shear rate domains.  Notably, it also predicts the narrowing intermediate window around $\phi = \phi_c$ in the limit of zero shear rate, as the quasi-static and inertial contributions to the stress become small near the jamming point.  The general form of the pressure model based on the Hookean-case results can hence be written as
\begin{align}
	p &= \left\{
     \begin{array}{ll}
       p_{\text{QS}} + p_{\text{Int}} & \mbox{for $\phi \ge \phi_c$}\\
       (p_{\text{Inert}}^{-1} +  p_{\text{Int}}^{-1})^{-1} & \mbox{for $\phi < \phi_c$}\\
     \end{array}
   \right.
   \label{eq:blend_p}
%   \\
%   \tau &= \left\{
%     \begin{array}{ll}
%       \tau_{\text{QS}} + \tau_{\text{Int}} & \mbox{for $\phi \ge \phi_c$}\\
%       (\tau_{\text{Inert}}^{-1} + \tau_{\text{Int}}^{-1})^{-1} & \mbox{for $\phi < \phi_c$}\\
%     \end{array}
%   \right.
%   \label{eq:blend_tau}
\end{align}
with the individual regime contributions defined as
\begin{align}
	p_{\text{QS}}d/k &= \alpha_{\text{QS}} |\phi - \phi_c|^{2/3}
%	&
%	\tau_{\text{QS}} &= \beta_{\text{QS}} |\phi - \phi_c|
	\label{eq:stress_QS}
	\\
	p_{\text{Int}}d/k &= \alpha_{\text{Int}} \hat{\dot{\gamma}}^{1/2}
%	&
%	\tau_{\text{Int}} &= \beta_{\text{Int}} \hat{\dot{\gamma}}^{5/7}
	\label{eq:stress_Int}
	\\
	p_{\text{Inert}}d/k &= \frac{\alpha_{\text{Inert}} \hat{\dot{\gamma}}^{2}}{|\phi - \phi_c|^{2}} \text{.}
%	&
%	\tau_{\text{Inert}} &= \frac{\beta_{\text{Inert}} \hat{\dot{\gamma}}^{2}}{|\phi - \phi_c|^{9/5}}
	\label{eq:stress_Inert}
\end{align}
The pressure at $\phi = \phi_c$ can be calculated using either blend, since Eqs.~\ref{eq:stress_QS} and \ref{eq:stress_Inert} yield $p_{\text{QS}}(\phi=\phi_c)=0$ and $p_{\text{Inert}}(\phi=\phi_c)=\infty$, which both yield $p = p_{\text{Int}}$ upon substitution into Eq.~\ref{eq:blend_p}; this case is included with the quasi-static blending solely for the sake of simplicity.  The constitutive parameter $\alpha_{\text{QS}}$ is a function of $\mu$, while $\alpha_{\text{Inert}}$ and $\alpha_{\text{Int}}$ are fairly $\mu$-independent.  These and other model constants are given in Table~\ref{tab:model}.

There are a few features of Eqs.~\ref{eq:blend_p}-~\ref{eq:stress_Inert} that are worth noting.  Firstly, for systems above the critical volume fraction, the blending function yields a model of Herschel-Bulkley form, which has been shown previously to capture the shear stress of soft-sphere systems~\cite{Nordstrom2010, Seth2008}.
Additionally, the individual regime contributions are consistent with some known scalings.  For example, the quasi-static pressure is proportional to the particle stiffness~\cite{Sun2011}, while $p_{\text{Inert}} = \alpha_{\text{Inert}} \rho_s (\dot{\gamma}d)^{2}/|\phi - \phi_c|^{2}$ rightly exhibits no dependence on $k$~\cite{Garzo1999, Lun1984, Jenkins1985, Johnson1987}.  Finally, the viability of the $\phi$-scaling in Eq.~\ref{eq:stress_Inert} for all $\mu$ values suggests that the $\phi_c = \phi_c(\mu)$ formulation could be a simple but effective step in improving current kinetic theory models.
\begin{table}
\caption{Values of model constants}
\centering
%\begin{ruledtabular}
\begin{tabular*}{\columnwidth}{@{\extracolsep{\fill}} lcccccccccc}
	\hline \hline
	\multicolumn{11}{l}{$\mu$-dependent parameters} \\
%	\hline
	$\mu$	&&	0.0	&&	0.1	&&	0.3	&&	0.5	&&	1.0	
	\\
	\hline
	$\eta_s$	&&	0.105	&&	0.268	&&	0.357	&&	0.382							&&	0.405	\\
	$\alpha_{\text{QS}}$		&&	0.095	&&	0.083	&&	0.14	&&	0.20							&&	0.25		\vspace{0.2cm}
\end{tabular*}
\begin{tabular*}{\columnwidth}{@{\extracolsep{\fill}} lcccccccccccccc}
%	\hline \hline
	\multicolumn{15}{l}{$\mu$-independent parameters} \\
%	\hline
	$\alpha_{\text{Inert}}$	&&	$\alpha_{\text{Int}}$	&&	$I_0$	&&	$\alpha_1$	&&	$\beta_1$	&&	$\hat{\dot{\gamma}}_0$	&&	$\alpha_2$	&&	$\beta_2$		\\
	\hline
	0.021	&&	0.099	&&	0.32	&&	0.37	&&	1.5	&&	0.1	&&	0.2	&&	1.0	\vspace{0.1cm}
	\\ \hline \hline
\end{tabular*}
%\end{ruledtabular}
\label{tab:model}
\end{table}

\section{Dimensionless groups and stress ratio model}
%\textit{Dimensionless groups and stress ratio model}.~---
It is possible to construct an analogous model for the shear stress as for the pressure, as previous works have shown $\tau$ to exhibit similar scalings with respect to the distance to jamming~\cite{Hatano2008,Otsuki2009b,Tighe2010}.  However, because $\tau$ and $p$ both vary over several orders of magnitude, fitting them directly can result in poor predictions of their ratio, $i.e.$ the shear stress ratio $\eta \equiv \tau/p$, which varies over a much narrower range.  For this reason, we choose to construct a model for $\eta$ and then express the shear stress as $\tau = \eta p$.

Some recent, successful rheological models for dense granular flows employ a dimensionless parameter called the inertial number as the basis for achieving stress collapses over a range of volume fractions and shear rates \cite{GDRmidi2004, Jop2006, daCruz2005}.
This inertial number $I \equiv \dot{\gamma}d/\sqrt{p/\rho_s}$ is a ratio of the timescales of shear deformation and particle rearrangement, and the physics of granular flows of hard particles is said to be determined by the competition of these two mechanisms.  When the particles have a finite stiffness, however, the binary collision time is nonzero and therefore presents yet another important timescale.  With this point in mind, we note that the dimensionless shear rate $\hat{\dot{\gamma}}$ identified earlier is in fact the ratio of the binary collision time to the macroscopic deformation time~\cite{Campbell2002}, and we show here that it can be used along with the inertial number to characterize soft particle rheology.

\begin{figure*}
	\centering
	\subfigure[]{
		\includegraphics[width=3.25 in]{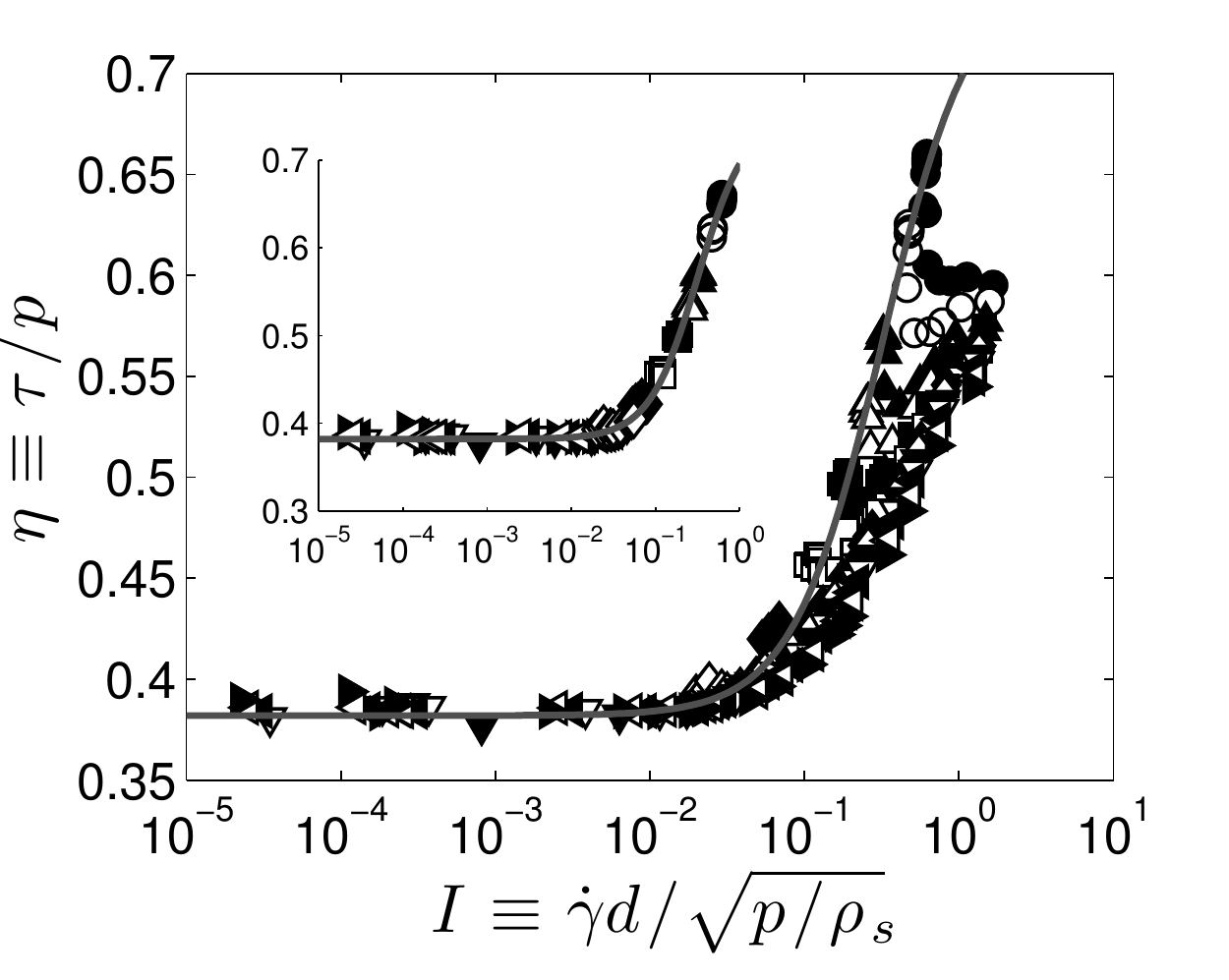}}
	\subfigure[]{
		\includegraphics[width=3.25 in]{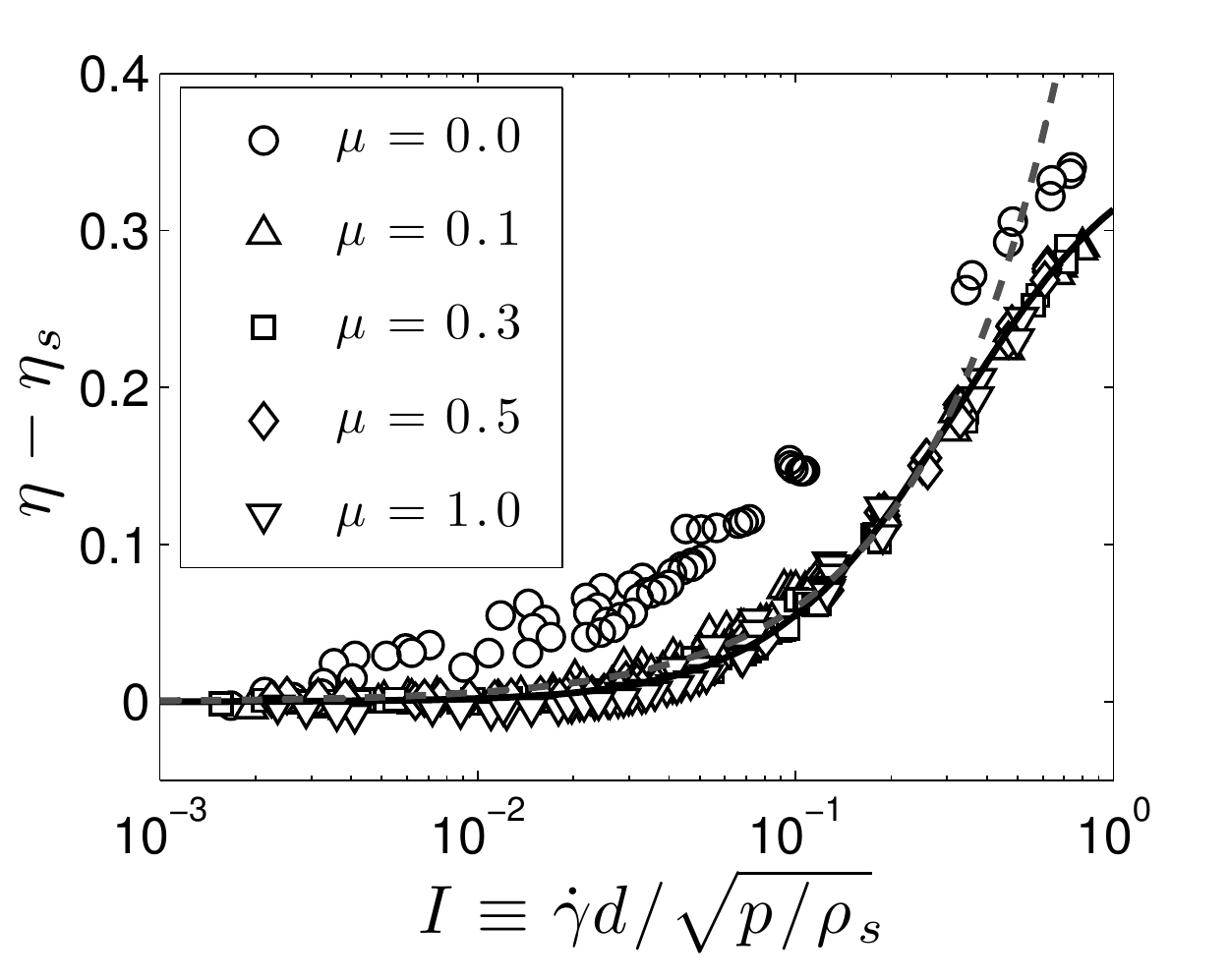}}
	\caption{Behavior of the shear stress ratio with respect to inertial number.
		(a) Significant scatter is observed when data from all three regimes are included.  Data are shown for $\mu = 0.5$ (see legend in Figure~\ref{fig:stress_gdot}a).  Inset:  A good collapse is achieved, however, for cases in which $\hat{\dot{\gamma}} \le 3.2\times 10^{-5}$. These cases correspond essentially to the quasi-static ($I \lesssim 10^{-2}$) and inertial regimes ($I \gtrsim 10^{-2}$) and are also indicated with a best-fit line (Eq.~\ref{eq:eta_hard}) in the main figure.  Intermediate regime data lie below this line.
		(b) The increase in the stress ratio from the yield stress ratio for these small-$\hat{\dot{\gamma}}$ cases collapses for $\mu \ge 0.1$.  Eq.~\ref{eq:eta_hard} captures these data well (solid line), as does the model of da Cruz et al.~\cite{daCruz2005} for $I \lesssim 0.3$ (dashed line).
	\label{fig:inertia1}}
\end{figure*}

In Figure~\ref{fig:inertia1} we plot the stress ratio versus the inertial number for $\mu = 0.5$.  For the densest systems, \emph{i.e.}~for low $I$, $\eta$ exhibits a constant-value asymptote that we identify as the yield stress ratio $\eta_s = \eta_s(\mu)$; values of $\eta_s$ for different cases of $\mu$ are presented in Table~\ref{tab:model}.  As $I$ increases, $\eta$ then also increases.
These same observations were made in previous studies of particles in the infinitely-hard limit~\cite{Jop2006, daCruz2005, GDRmidi2004}.  However, unlike in these works, we also observe significant scatter as $I$ becomes larger, which we will now show to be a consequence of the particle softness.

Because the inertial number models are designed for hard particles, we first limit our analysis to cases in which particle softness has little effect, i.e.~for small $\hat{\dot{\gamma}}$.  Indeed, quasi-static and inertial regime data of $\eta$ versus $I$ from our DEM simulations collapse onto a single curve, with the quasi-static regime occurring for $I \lesssim 10^{-2}$ and inertial regimes occurring for $I \gtrsim 10^{-2}$.  This collapse is seen in the inset of Figure~\ref{fig:inertia1}a for $\mu = 0.5$.  We model this curve as
\begin{align}
	\eta_{\text{hard}}(I) = \eta_s(\mu) + \frac{\alpha_1}{(I_0/I)^{\beta_1}+1}\text{,}
%	\eta_{\text{hard}}(I) = \eta_s + \alpha_1 I^{\beta_1} \text{,}
	\label{eq:eta_hard}
\end{align}
where $I_0$, $\alpha_1$, and $\beta_1$ are parameters dictating the transition from quasi-static to inertial flow.  This form is similar to that of Jop \emph{et al.}~\cite{Jop2006}.
%Jop and coworkers~\cite{Jop2006}.  We choose $\eta_{\infty}$ to agree with the dilute limit of $\eta$ from a kinetic theory of inelastic spheres~\cite{Garzo1999} whose effective restitution coefficient changes with $\mu$~\cite{Jenkins2002}; for $\mu = 0.5$ and $e = 0.7$, we set $\eta_{\infty} = 0.84$.
%As seen in Table~\ref{tab:model}, the values of the fitting parameters change very little for $\mu \ge 0.1$.  Since the interparticle friction coefficient for most real granular materials falls in this range, we take $\alpha_1 = 0.44$ and $\beta_1 = 0.8$ as as suitable averages for our model.  With the same values of $\alpha_1$ and $\beta_1$, it is clear that $\eta_{\text{hard}}(I) - \eta_s$ will collapse for all of these $\mu$ cases (see Figure~\ref{fig:inertia1}b).
Interestingly, the increase of $\eta$ from $\eta_s$ is nearly identical for all cases of $\mu \ge 0.1$ (Fig.~\ref{fig:inertia1}b).  Since the interparticle friction coefficient for most real granular materials falls in this range, we conveniently take one set of constitutive parameter values as suitable averages for our model; these values are presented in Table~\ref{tab:model}.
%, with $I_0 = 0.32$, $\alpha_1 = 0.37$, and $\beta_1 = 1.5$.

%\begin{table}
%\caption{Values of rheological model constants.}
%%\caption{Parameter values for the shear stress ratio model.  For $\mu \ge 0.1$, we can approximate $\alpha_1 = 0.44$, $\beta_1 = 0.8$, $\alpha_2 = 0.86$, and $\beta_2 = 0.80$.}
%\centering
%\begin{ruledtabular}
%\begin{tabular}{lccccccccccc}
%%	\textbf{$\mu$}		&&	0.0	&&	0.1	&&	0.3	&&	0.5	&&	1.0	\\
%\omit \text{Pressure model constants}	&\vspace{0.2cm}	&&&&&&&&&& \\
%\hline
%\omit \text{Shear stress ratio model constants}	&	&&&&&&&&&& \\
%\hline
%		&	$I_0$		&&	$\alpha_1$	&&	$\beta_1$	&&	$\hat{\dot{\gamma}}_0$		&&	$\alpha_2$	&&	$\beta_2$		\\
%	\hline
%%	\textbf{$m^*$}	&&	0.475	&&	0.550	&&	0.529	&&	0.495							&&	0.441	
%	
%%	\\
%%	\textbf{$\alpha_1$}	&&	0.499	&&	0.495	&&	0.561	&&	0.591							&&	0.715	
%%	\\
%%	\textbf{$\beta_1$}	&&	0.517	&&	0.811	&&	0.901	&&	0.868							&&	1.01	
%%	\\
%%	\textbf{$\alpha_2$}	&&	0.626	&&	0.980	&&	1.20	&&	1.15							&&	1.71	
%%	\\
%%	\textbf{$\beta_2$}	&&	0.727	&&	0.931	&&	0.920	&&	0.850							&&	0.983	
%\end{tabular}
%\end{ruledtabular}
%\label{tab:model}
%\end{table}

The form of $\eta_{\text{hard}}$ presented in Eq.~\ref{eq:eta_hard} is not the only viable option.  Another possibility is a simple power law, which can be written as
$\eta_{\text{hard}}(I) = \eta_s + \alpha_1^\prime I^{\beta_1^\prime}$.  This form has been used previously by da Cruz and coworkers~\cite{daCruz2005} with $\beta_1^\prime = 1$.  A comparison between this form, with $\beta_1^\prime = 1$ and $\alpha_1^\prime = 0.6$, and the one in Eq.~\ref{eq:eta_hard} are shown in Figure~\ref{fig:inertia1}b.  The two models agree closely for all values of $I \lesssim 0.3$, with a departure occurring for larger $I$.  However, with the inertial number models, we need to be concerned only with volume fractions greater than the freezing transition $\phi_f = 0.49$~\cite{Torquato1995}, where traditional kinetic theories fail~\cite{Jenkins2010, Forterre2008}.  At $\phi_f$, the kinetic theory of Garz\`o and Dufty~\cite{Garzo1999} predicts $I = 0.83$, which is consistent with our DEM results and beyond which we can ignore disparities in the $\eta_{\text{hard}}$ predictions between the two models.  Hence, though we continue with Eq.~\ref{eq:eta_hard}, we view both forms as being acceptable.

%However, inertial numbers beyond about 0.8 (which corresponds to the freezing point $\phi_f = 0.49$~\cite{Torquato1995}) 

%inertial number models are intended for flows denser than the freezing transition $\phi_f = 0.49$~\cite{Torquato1995}, with 
%These results are included in Figure~\ref{fig:inertia1} for comparison with those of Eq.~\ref{eq:eta_hard}.  The two models agree closely for all values of $I$ except .

\begin{figure*}
	\centering
	\subfigure[]{
		\includegraphics[width=3.25 in]{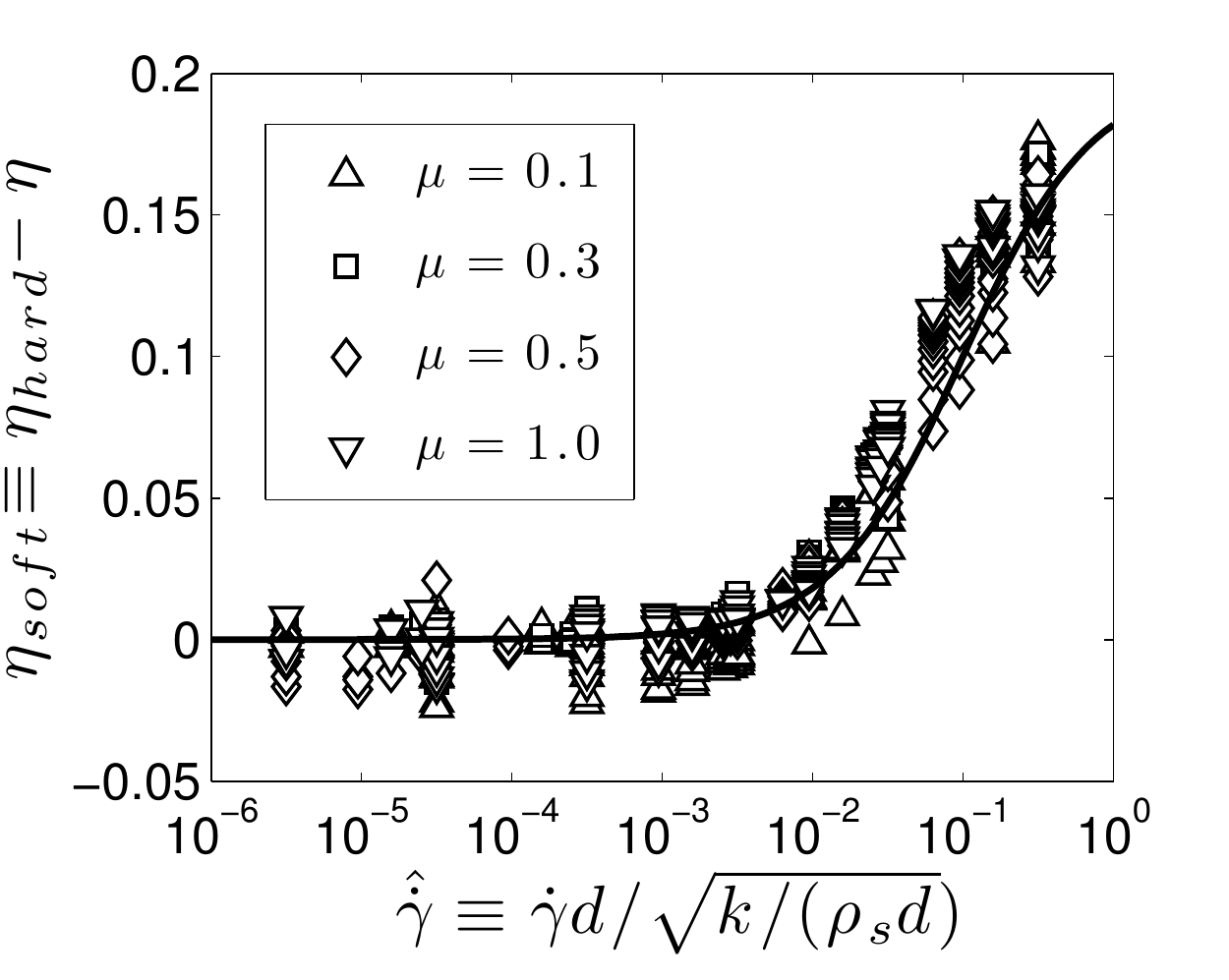}}
	\subfigure[]{
		\includegraphics[width=3.25 in]{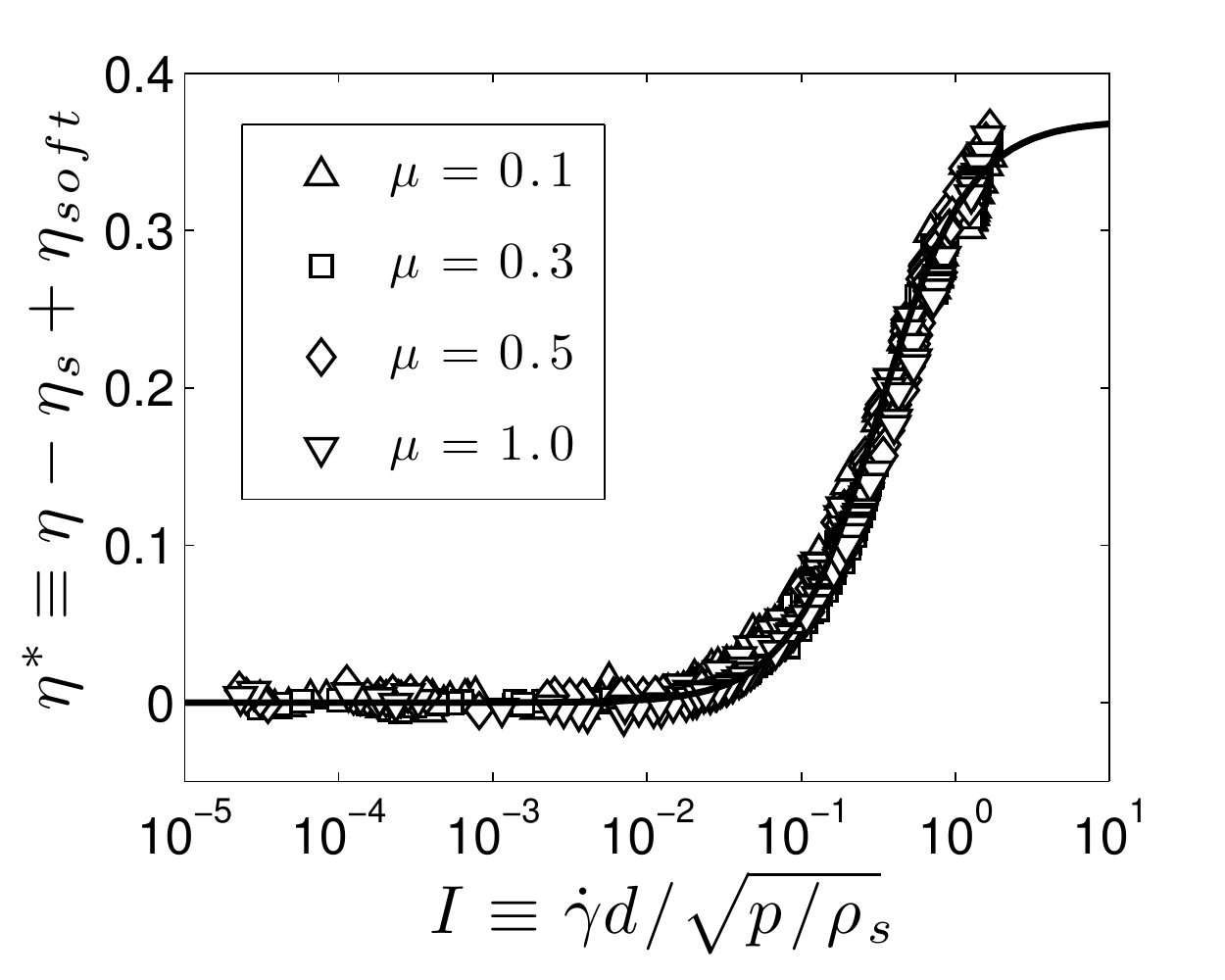}}
	\caption{Shear stress ratio contribution from $\hat{\dot{\gamma}}$ for all values of $\hat{\dot{\gamma}}$ and $\mu \ge 0.1$.
	(a) The softness-induced departure $\eta_{\text{soft}}$ of the stress ratio from its hard-particle limit is essentially a function of only $\hat{\dot{\gamma}}$.
	(b) The correction for particle softness yields a collapse of the data in all three regimes.
	\label{fig:inertia2}}
\end{figure*}

Though Eq.~\ref{eq:eta_hard} captures low-$\hat{\dot{\gamma}}$ behavior well, inclusion of higher-$\hat{\dot{\gamma}}$ cases reveals a noticeable departure from the $\eta_{\text{hard}}(I)$ curve, as seen in Figure~\ref{fig:inertia1}.  Specifically, for a given value of $I$, the value of $\eta$ from an intermediate-regime flow is consistently lower than that given by the Eq.~\ref{eq:eta_hard}.
This deviation is a consequence of particle softness and, in the context of our regime blending, grows in magnitude with the intermediate-regime contribution to the pressure.  Figure~\ref{fig:inertia2}a shows the connection between the magnitude of this departure $\eta_{\text{soft}} \equiv \eta_{\text{hard}} - \eta$ and $\hat{\dot{\gamma}}$.  This softness effect, similarly to $\eta_{\text{hard}}$, can be modeled as
\begin{align}
%	\eta_{\text{soft}}(\hat{\dot{\gamma}}) = \alpha_2 \hat{\dot{\gamma}}^{\beta_2} \text{,}
	\eta_{\text{soft}}(\hat{\dot{\gamma}}) = \frac{\alpha_2}{(\hat{\dot{\gamma}}_0/\hat{\dot{\gamma}})^{\beta_2} + 1} \text{,}
	\label{eq:eta_soft}
\end{align}
where $\hat{\dot{\gamma}}_0$ = 0.1, $\alpha_2 = 0.2$, and $\beta_2 = 1$ are constants describing the transition to intermediate flow.
%We note here that, interestingly, $\beta_1 \approx \beta_2$.
Finally, we can write
\begin{align}
	\eta(I, \hat{\dot{\gamma}}) = \eta_{\text{hard}}(I) - \eta_{\text{soft}}(\hat{\dot{\gamma}}) \text{,}
	\label{eq:eta_final}
\end{align}
and, by plotting $\eta^* \equiv \eta - \eta_s + \eta_{\text{soft}}$ vs.~$I$ as in Figure~\ref{fig:inertia2}b, we arrive at a collapse of the stress ratio data from all three regimes.

%One important observation is that data in the quasi-static regime lie on the $\eta_{\text{hard}}(I)$ curve despite the fact that particles in this regime exhibit non-negligible particle overlaps.  While it is true 

%Though infinitely hard particles can exist only in the inertial regime, we have observed that quasi-static regime data also lie on the .

\section{Generalized continuum model}

Our rheological model therefore consists of Eqs.~\ref{eq:blend_p} - \ref{eq:stress_Inert} for the pressure and Eqs.~\ref{eq:eta_hard} - \ref{eq:eta_final} for the shear stress ratio.  Though the collapses can generally be improved by allowing the fitting parameters to be functions of $\mu$ rather than constants, the fits are nevertheless fairly good and hence justify the use of simpler forms.  

While this model was developed for simple shear flows, it can be recast to handle general deformation types as done in Ref.~\cite{Sun2011}.  First, we note that the strain rate tensor for simple shear flows is
$\mathbf{D} = \frac{1}{2}\dot{\gamma}(\mathbf{e}_x\mathbf{e}_z + \mathbf{e}_z\mathbf{e}_x)$
where $\mathbf{e}_i$ are the unit vectors in the $i$ direction.  This expression can be rearranged to yield
\begin{align}
	\dot{\gamma} = 2 | \mathbf{D} | \text{,}% \sqrt{2 \mathbf{D}:\mathbf{D}} \text{,}
	\label{eq:gdot_general}
\end{align}
where $|\mathbf{D}| = \sqrt{\frac{1}{2}\mathbf{D^T}:\mathbf{D}}$ is the modulus of $\mathbf{D}$, and $\mathbf{D}$ is taken to correspond to general deformation types.  Finally, we write the stress tensor as 
\begin{align}
	\boldsymbol{\sigma} = p \hspace{0.05cm} (\mathbf{I}-\eta \hat{\mathbf{S}})
	\label{eq:sigma_model}
\end{align}
where $p$ and $\eta$ are given by our model, $\mathbf{I}$ is the identity tensor, and $\hat{\mathbf{S}} = \mathbf{S}/|\mathbf{D}|$ with $\mathbf{S}=\mathbf{D}-\frac{1}{3}\text{tr}(\mathbf{D})$.  Eqs.~\ref{eq:gdot_general} and~\ref{eq:sigma_model} allow our rheological model to handle flows in more complex geometries as are commonly found in real flow scenarios.

\section{Summary}
%\textit{Summary}.~---
We have investigated shear flows of dense frictional granular materials in all three flow regimes in order to gain a better understanding of the scalings within each regime and the transitions between them.  We find scaling relations for the pressure with respect to both shear rate and the distance to the jamming point and, for the intermediate regime, observe identical power-law behavior for particles with different friction coefficients.  Furthermore, we propose a simple blending function for patching each regime's asymptotic form in order to predict pressure in between regimes.  Finally, we decompose the shear stress ratio into contributions from two dimensionless shear rates, enabling us to quantify the effect of particle softness.  These findings establish a framework for a global model for steady-state simple shear flows of dense granular matter.

We gratefully acknowledge the support of DOE/NETL Grant No. DE-FG26-07NT43070.

%\bibliographystyle{apsrev}
%\bibliography{PRL_schialvo_2011}

\end{document}